**Single Iteration Conditional Based DSSE Considering Spatial and Temporal Correlation**

Mehdi Shafiei[a], Ghavameddin Nourbakhsh[a,*], Ali Arefi[b], Gerard Ledwich[a], Houman Pezeshki[a]

[a]Department of Electrical and Electronic Engineering and Computer Science, Queensland University of Technology, Australia

[b]School of Engineering and Information Technology, Murdoch University, Australia

## Abstract

The increasing complexity of distribution networks calls for advancement in distribution system state estimation (DSSE) to monitor the operating conditions more accurately. A sufficient number of measurement devices is imperative for a reliable and accurate state estimation. The limitation on the measurement devices is generally tackled with using the so-called pseudo measured data. However, the errors in pseudo data by current techniques are quite high leading to a poor DSSE. As customer loads in distribution networks show high cross-correlation in various locations and over successive time steps, it is plausible that deploying the spatial-temporal dependencies can improve the pseudo data accuracy and estimation. Although the role of spatial dependencies in DSSE has been addressed in the literature, one can hardly find an efficient DSSE framework capable of incorporating temporal dependencies present in customer loads. Consequently, to obtain a more efficient and accurate state estimation, we propose a new non-iterative DSSE framework to involve spatial-temporal dependencies together. The spatial-temporal dependencies are modeled by conditional multivariate complex Gaussian distributions and are studied for both static and real-time state estimations, where information at preceding time steps are employed to increase the accuracy of DSSE. The efficiency of the proposed approach is verified based on the quality and accuracy of the indices, standard deviation and computational time. The method applied to a combination of residential and industrial customer loads in three different balanced medium voltage (MV), and one unbalanced low voltage (LV) distribution case studies for evaluations.

## 1. Introduction

Power system state estimation (SE) has been used extensively for transmission systems operation and control since its first introduction in 1970 [1]. Unlike transmission networks, SE is not commonly applied to distribution networks due to complexities with limited real measurements available in distribution networks. However, with increasing penetration of distributed energy resources (DERs) like rooftop PVs and electric storage systems in distribution networks, an efficient and practical form of distribution system state estimation (DSSE) is indispensable [2]. The early development of DSSE goes back to 1990 when the weighted least square (WLS) algorithm was designed and applied to a distribution system [3]. Later in 1996, a DSSE was designed and employed as a real-time monitoring in distribution management system (DMS) for applications such as volt/var control considering the impacts of DERs, feeder reconfiguration, battery storage management, and protection [4, 5]. Multiphase DSSE approaches suitable for LV networks in the presence of DERs are proposed in [6].

A high penetration of DERs in a distribution network on one hand, and unpredictable customer loads behaviour, on the other hand, require fast and accurate DSSE methods for online generation/load demand considerations and planning [7, 8]. However, a high number of customer loads and a huge amount of measured data from smart meters make centralized DSSE algorithms complicated and computationally demanding [9]. An enhanced form of DSSE with a significant reduction in measurement points, while retaining accurate estimations, can be an attractive alternative. Generally, decreasing the number of measurement points leads to an under-determined system, meaning that the measurements cannot provide sufficient information required for an accurate state estimation algorithm [10]. This problem is resolved using pseudo measurements for unmeasured buses, satisfying the distribution network observability conditions [11]. Pseudo measurements can be obtained from advanced metering infrastructure (AMI), historical data or customers' billing data [12], or they can be calculated based on the nominal customers' load power, consumed power and daily load profiles [13]. Although the use of pseudo measurement has become an essential part of the DSSE algorithms, the associated error with this type of measurement data is still substantial, leading to DSSE with low accuracy and reliability. Spatial correlation information between real measured and unmeasured points is deployed to improve DSSE accuracy in [14]. In [15], the spatial correlation between loads is used to further increase the accuracy of the estimated injected currents in unmeasured buses. Similarly, in [16], spatial dependencies are modeled to determine the pseudo load profile for unmeasured customers where the essential load patterns are extracted from the smart meters data using clustering techniques.

Although the impact of incorporating spatial correlation information on the accuracy of DSSE is addressed in the literature, there is no developed DSSE formulation capable of including temporal dependency or spatial-temporal dependencies. This is while customer loads, in general, show high correlations in successive time steps. The temporal correlation can offer a measure for the similarity of load(s) variations in time, greatly enhancing estimation quality. Recently, in the forecasting literature, the great potential of involving spatial-temporal dependencies in improving PV power prediction performance has been verified [17]. Also, spatial-temporal correlations are considered for load growth forecasting and load demand in electrical vehicles (EVs) charging patterns [18]. A Vector Auto-Regressive (VAR) model is considered in [19] to integrate time and space correlations present in measured data into a DSSE algorithm. In this article, WLS as an iterative algorithm is employed for state estimation in the presence of several phasor measurement units. The iterative based algorithms with the large amount of required data in a distribution network make these estimation processes computationally time-consuming, while an active distribution network with DERs requires fast and accurate state estimators, updating data in less than one second [5]. To deal with a large amount of measured data, a new algorithm based on compressed measurements is proposed in [20]. However, the iterative DSSE algorithm proposed in [20] computationally is highly demanding.

This work aims at developing a computationally efficient and accurate DSSE framework to incorporate spatial-temporal dependencies of customer loads. The proposed approach is a single iteration technique suitable for real-time DSSE using pseudo measurement data. We study the spatial-temporal correlation strength of customer loads as a function of a number of customers in each residential region. The correlation strength of net load with 15% to 25% PV power penetration is also discussed. The presence of high spatial-temporal correlations between net loads in distribution networks motivates us to incorporate such dependency information in the proposed DSSE formulation. Conditional multivariate complex Gaussian distribution (CMCGD) is used to characterize spatial-temporal correlations. With the focus on temporal correlations, in the proposed formulation, the measured data in the previous time steps are integrated to decrease the error of estimation. This results in a significant reduction of pseudo measurements errors. The proposed approach is computationally very efficient, making it suitable for real-time state estimation. The performance of the proposed method is evaluated based on four case studies including an unbalance LV distribution networks.

## 2. Spatial and Temporal Correlation

Correlation is a statistical relationship between two random variables. Spatial correlation is computed based on the data from different locations, while temporal correlation represents the degree of similarities between data in different time steps [21]. The temporal and spatial correlation coefficients are usually defined mathematically as in (1) [22]:

$$\begin{cases} Correlation_{Spatial} = {cov(\boldsymbol{d1}, \boldsymbol{d2})}/{\sigma_{\boldsymbol{d1}} * \sigma_{\boldsymbol{d2}}} \\ Correlation_{Temporal} = {cov(\boldsymbol{d1}_t, \boldsymbol{d1}_{t+k})}/{var(\boldsymbol{d1})} \end{cases} \tag{1}$$

where $cov$, $\sigma$ and, $var$ denote covariance, standard deviation and variance, respectively. For spatial correlation, $\boldsymbol{d1}$ and $\boldsymbol{d2}$ represent two sets of data from two different geographic locations, while temporal correlation describes the dependency at a given location and between time intervals of $t$ and $t + k$.

In this section, the aim is to study the spatial-temporal correlation strength between net loads in a distribution system based on two important factors:

- The number of loads in customer residential communities,
- Presence of DERs at load buses.

In order to study the first noted factor, we consider two residential communities (RC1 and RC2) connected to the similar distribution transformers, while the number of RC customers is gradually increased stepwise, as shown in Figure 1. The aggregated consumed power in RC1 and RC2 in each step generate $\boldsymbol{d1}$ and $\boldsymbol{d2}$ in (1) to calculate spatial correlation. Furthermore, $\boldsymbol{d1}$ is employed to find the temporal correlation of each step as well.

The spatial and temporal correlation coefficients are computed and shown in Figure 2. For the analysis, we use one-minute active power data of two real datasets from the Newmarket suburb in Brisbane, Australia in the summer season, and Pecan Street, Texas, USA in winter season [23].

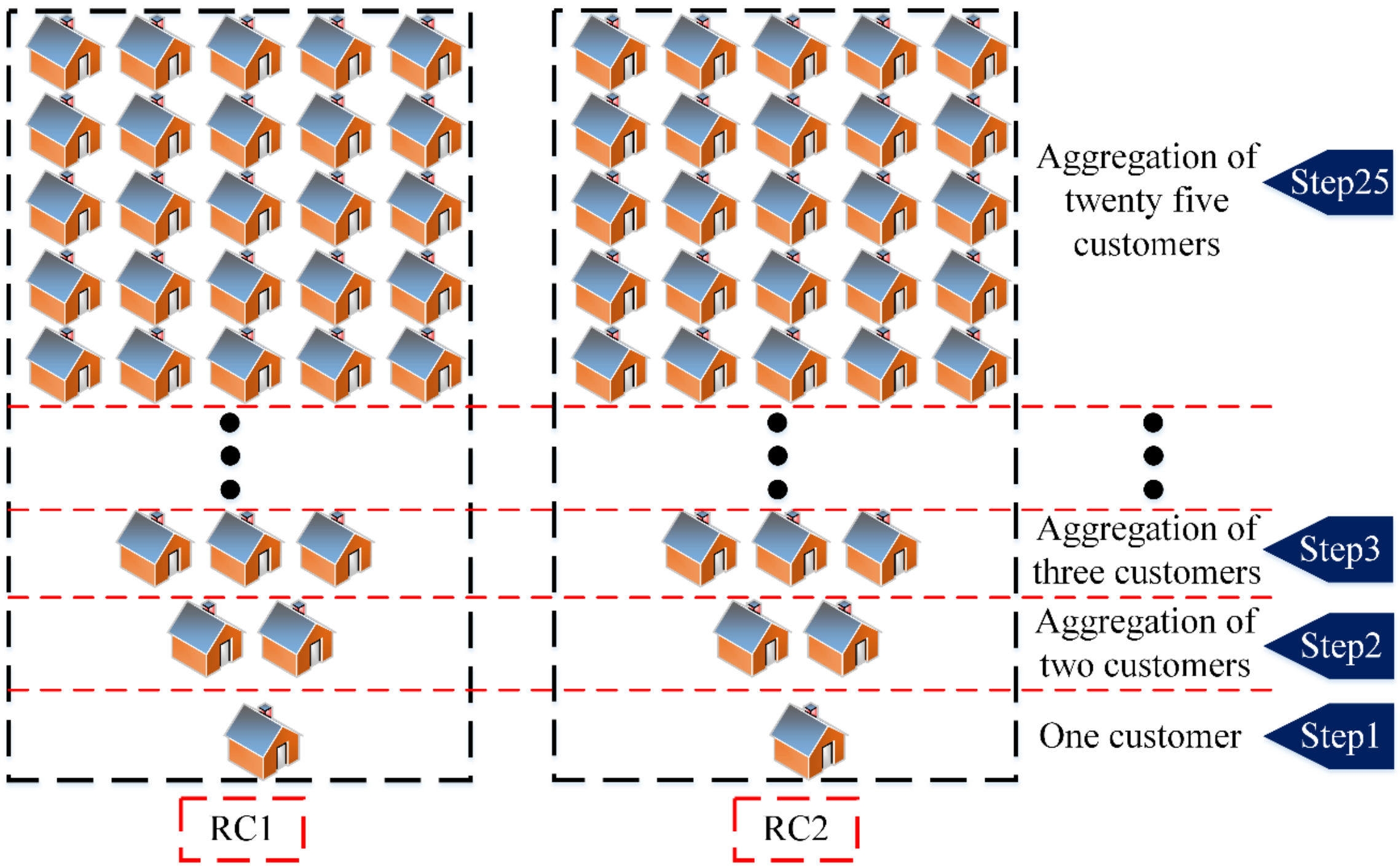

Figure. 1. Gradual increase in the number of household customers from step 1 – 25, in each RC.

The spatial correlations are calculated for the aggregated loads in each community over the course of 7 days. The temporal correlation coefficients represent the correlation strength between successive time steps of the aggregated loads. According to Figure 2, as the number of customers increases, the spatial correlation between the aggregated loads in two RCs increases. This happens due to smoothing effects of aggregation. As one observes in Figure 2 (a) and Figure 2 (b), the spatial correlation coefficient between two individual houses is less than 20% and it increases to more than 80% when the number of houses in each RC rises to 25. The temporal correlation in Figure 2 (c) and Figure 2 (d) shows a similar trend. Note that; in Figure 2 (c), the irregular jump at the case with two houses is accidental and does not imply to be a general trend.

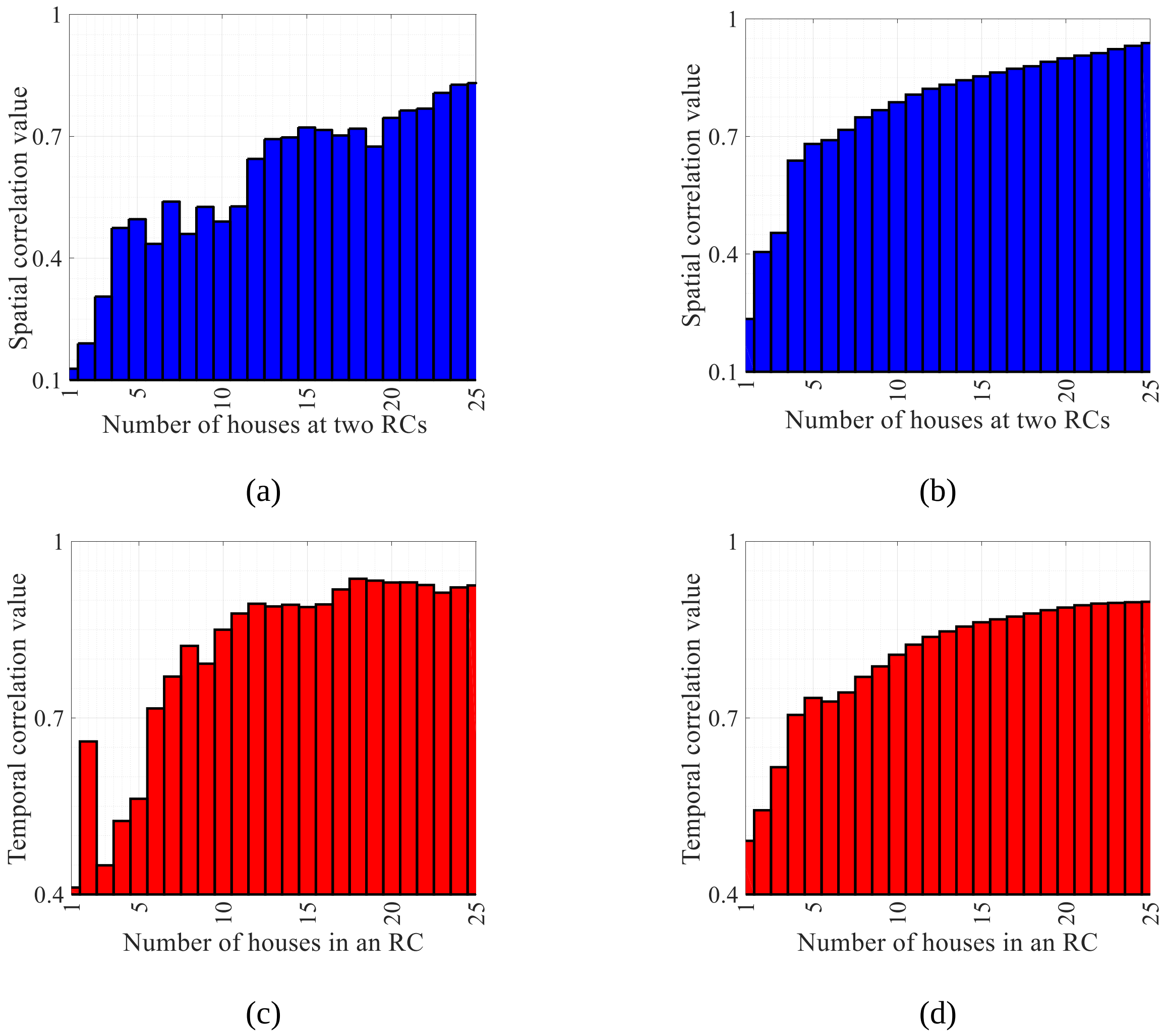

Figure. 2. Impact of customer numbers on spatial correlation (a) Newmarket, (b) Pecan street, and on temporal correlation (c) Newmarket, (d) Pecan street.

Furthermore, load aggregation decreases the variability in the customers' load profiles. Figure 3 shows a one-day load profile with one minute time samples for one customer (a), and a group of twenty five customers (b). Based on the results of this section, it is worth noting that load aggregation in MV/LV feeders or different segments of the LV distribution networks increases the spatial-temporal correlation, and decreases the variability in the load profiles. This high spatial-temporal correlation with a lower variability in load profiles considerably increases the accuracy of the CMCGD method in updating the unmeasured distribution nodes.

To investigate the second factor, five residential areas Area 1 (A1), Area 2 (A2), Area 3 (A3), Area 4 (A4), and Area 5 (A5) with different load levels and PV penetration rates (between 15% to 25%) from the Newmarket suburb data is used. Figure 4 (a) illustrates the correlation coefficients between PV power outputs. It is expected to observe high correlations between nearby PV sources within a distribution network [17]. This is also confirmed and indicated by unreadable dark colours in Figure 4 (a) representing correlation coefficients of higher than 0.95 between PV outputs in five areas, while correlation coefficient is nearly 0.7 for customer loads, as shown in Figure 4 (b). Interestingly, as depicted in Figure. 4 (c), the correlation rates of 'net loads' is very close to that of pure 'customer loads' in a close neighbourhood. The relevance of the above analysis in our proposed formulation is threefold, as described in the followings. (1) The proposed method is developed based on the spatial-temporal correlation strength between loads. A higher correlation rate leads to better estimation and accuracy in the proposed DSSE. Interestingly, by aggregation of houses in each RC, it is expected to see higher correlation rates. (2) As shown in Figure 1, the rate of increase in spatial-temporal correlation of loads at two different countries and in different seasons but with the same load type is very similar in values and trends. Hence, for new-built areas or those without historical data, the correlation matrix from areas with the same load type can be considered in the proposed DSSE formulation.

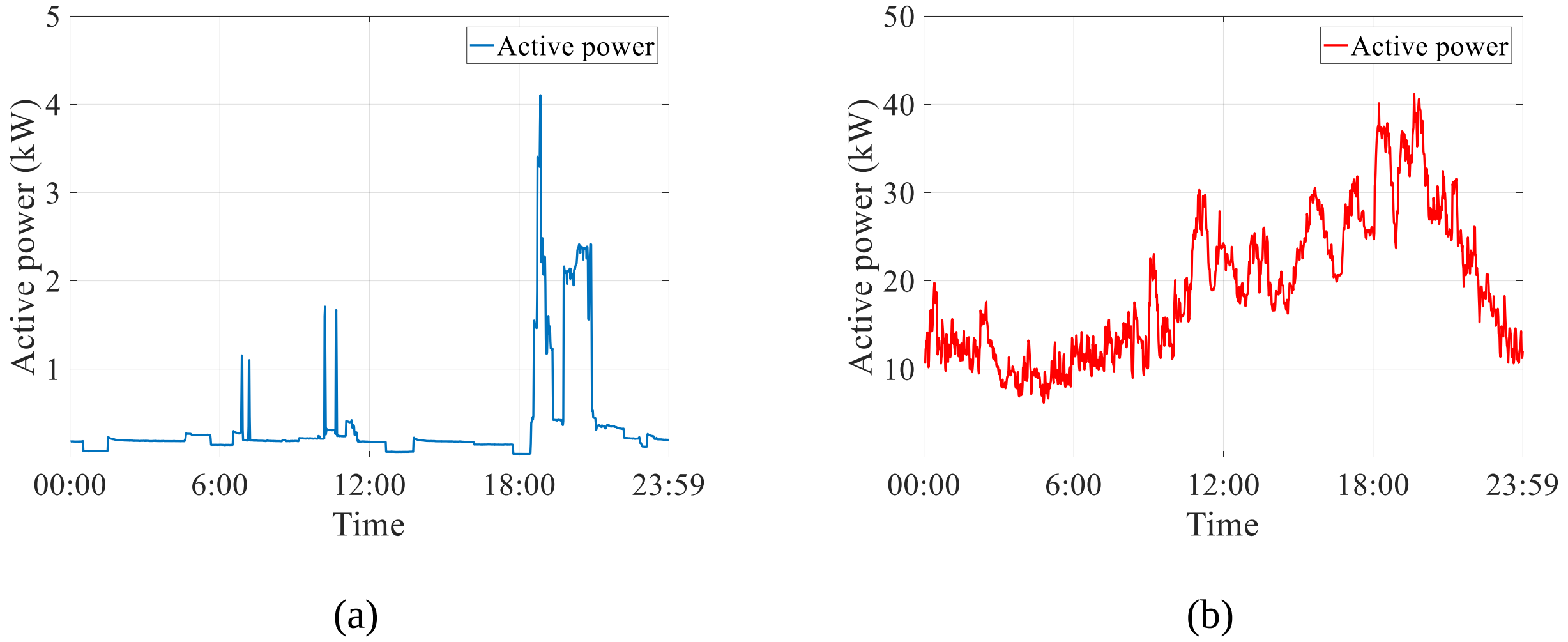

Figure 3. One day load profile (a) One customer, (b) Twenty customers.

(3) The PV penetration rate from zero to 25% does not change the correlation coefficients of the net load, noticeably. Therefore, in case the historical data for ever-growing PV sources is not available, one can rely on the latest correlation information available.

## 3. Spatial-Temporal Distribution State Estimation: Basics and Formulation

The aim of designing the proposed DSSE is to improve the accuracy of the pseudo data, based on spatial-temporal dependencies to estimate the states in a single iteration. In this paper, CMCGD with a built-in spatial-temporal correlation framework is proposed for DSSE.

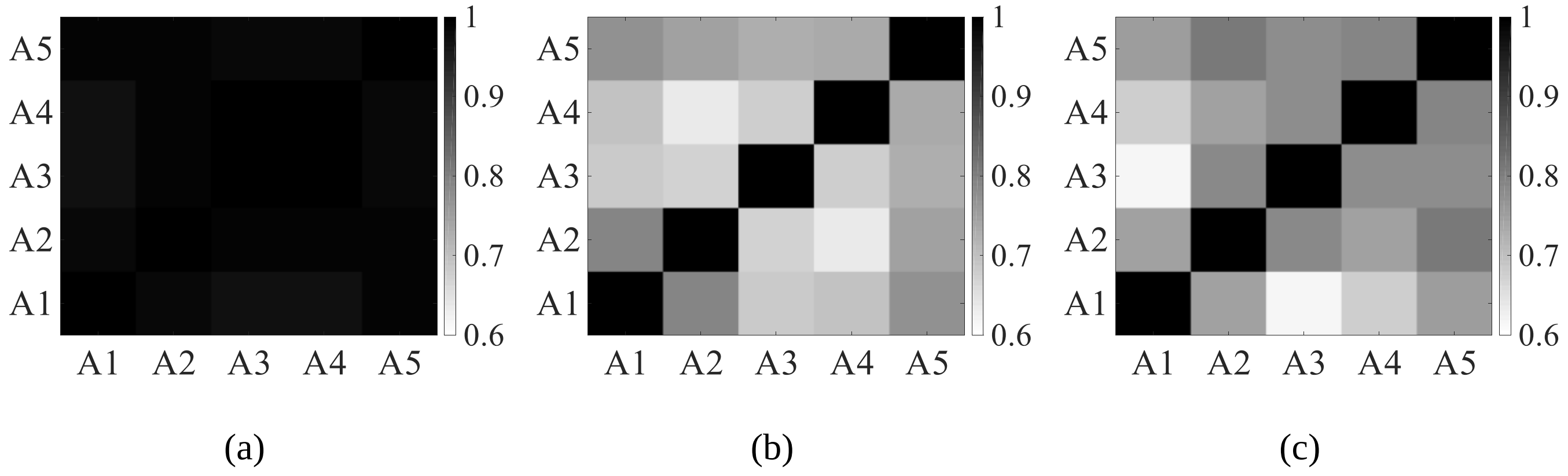

Figure. 4. Visualization of the correlation matrix for five groups of loads and PV generations, 5×5 matrix, and black to white colours represent the highest to lowest correlation; (a) PV outputs correlation, (b) customer loads correlation, (c) net loads correlation.

### 3.1 Conditional Multivariate Complex Gaussian Distribution

Denote ${\boldsymbol{y}_1}^{\boldsymbol{measured}}$ as the set of the measured data and ${\boldsymbol{y}_2}^{\boldsymbol{umeasured}}$ as the set of unmeasured data with mean values $(\boldsymbol{\mu}_1, \boldsymbol{\mu}_2)$, covariance matrix $(\boldsymbol{\Gamma}_{11}, \boldsymbol{\Gamma}_{22})$ and pseudo covariance $(\boldsymbol{C}_{11}, \boldsymbol{C}_{22})$ matrices, respectively. Multivariate Complex Gaussian Distribution (MCGD) with the mean matrix $(\boldsymbol{\mu})$, covariance matrix $(\boldsymbol{\Gamma})$ and pseudo covariance matrix $(\boldsymbol{C})$ describing the joint dependency of ${\boldsymbol{y}_1}^{\boldsymbol{measured}}$ and ${\boldsymbol{y}_2}^{\boldsymbol{umeasured}}$ is formulated as [24]:

$$\boldsymbol{\Gamma} = \begin{bmatrix} \boldsymbol{\Gamma}_{11} & \boldsymbol{\Gamma}_{12} \\ \boldsymbol{\Gamma}_{21} & \boldsymbol{\Gamma}_{22} \end{bmatrix}, \boldsymbol{C} = \begin{bmatrix} \boldsymbol{C}_{11} & \boldsymbol{C}_{12} \\ \boldsymbol{C}_{21} & \boldsymbol{C}_{22} \end{bmatrix} \tag{2}$$

where $\boldsymbol{\Gamma}_{12}$ and $\boldsymbol{\Gamma}_{21}$ are equal covariance matrices whose elements are the covariance between elements of ${\boldsymbol{y}_1}^{\boldsymbol{measured}}$ and ${\boldsymbol{y}_2}^{\boldsymbol{umeasured}}$. Similarly, $\boldsymbol{C}_{12}$ and $\boldsymbol{C}_{21}$ are equal pseudo covariance matrices whose elements are the pseudo covariance between elements of ${\boldsymbol{y}_1}^{\boldsymbol{measured}}$and ${\boldsymbol{y}_2}^{\boldsymbol{umeasured}}$. With $[\boldsymbol{\mu}_1, \boldsymbol{\Gamma}_{11}, \boldsymbol{\Gamma}_{12}, \boldsymbol{C}_{11}, \boldsymbol{C}_{12}]$ available, $[\boldsymbol{\mu}_2, \boldsymbol{\Gamma}_{22}, \boldsymbol{C}_{22}]$ can be updated as:

$$\begin{aligned} {\boldsymbol{\mu}_2}^{\boldsymbol{updated}} &= \boldsymbol{\mu}_2 + \boldsymbol{A}\left(\boldsymbol{\mu}_1^{\boldsymbol{M}} - \boldsymbol{\mu}_1\right) + \boldsymbol{B}(\boldsymbol{\mu}_1^{\boldsymbol{M}} - \boldsymbol{\mu}_1)^* \\ {\boldsymbol{\Gamma}_{22}}^{\boldsymbol{updated}} &= \boldsymbol{\Gamma}_{22} - \boldsymbol{A}\boldsymbol{\Gamma}_{12}^{\boldsymbol{H}} - \boldsymbol{B}\boldsymbol{\Gamma}_{12}^{\boldsymbol{H}} \\ {\boldsymbol{C}_{22}}^{\boldsymbol{updated}} &= \boldsymbol{C}_{22} - \boldsymbol{A}\boldsymbol{C}_{12}^{\boldsymbol{T}} - \boldsymbol{B}\boldsymbol{C}_{12}^{\boldsymbol{T}} \end{aligned} \tag{3}$$

$$\{\, \boldsymbol{A} = \left(\boldsymbol{\Gamma}_{12} - \boldsymbol{C}_{12}\boldsymbol{\Gamma}_{22}^{-*}\boldsymbol{C}_{22}^{\boldsymbol{H}}\right)\boldsymbol{\Lambda}_{22}^{-*},\ \boldsymbol{B} = \left(\boldsymbol{C}_{12} - \boldsymbol{\Gamma}_{12}\boldsymbol{\Gamma}_{22}^{-1}\boldsymbol{C}_{22}\right)\boldsymbol{\Lambda}_{22}^{-1},\ \boldsymbol{\Lambda}_{22} = \boldsymbol{\Gamma}_{22}^{*} - \boldsymbol{C}_{22}^{\boldsymbol{H}}\boldsymbol{\Gamma}_{22}^{-1}\boldsymbol{C}_{22} \}$$

where $(.)^*$ shows the complex conjugate operation, $(.)^{-1}$ is an inverse operation, $(.)^{-*}$ means inverse and conjugation operation, vector $\boldsymbol{\mu}_1^{\boldsymbol{M}}$ is the measured $\boldsymbol{\mu}_1$ and ${\boldsymbol{\mu}_2}^{\boldsymbol{updated}}$, ${\boldsymbol{\Gamma}_{22}}^{\boldsymbol{updated}}$ and ${\boldsymbol{C}_{22}}^{\boldsymbol{updated}}$ are the updated mean, covariance and pseudo covariance matrices of the unmeasured data. $\boldsymbol{A}$ and $\boldsymbol{B}$ are functions of covariance and pseudo covariance matrices as indicated in [25]. Equation (3) is a direct calculation that updates mean, covariance and pseudo covariance of the unmeasured data $({\boldsymbol{y}_2}^{\boldsymbol{umeasured}})$, by importing a new measured data $({\boldsymbol{y}_1}^{\boldsymbol{measured}})$.

It is worth noting that in cases where the measured data from distribution networks with high integration of DERs cannot be represented by a single Gaussian distribution, a mixture of Gaussian models can be used to characterize the stochastic behaviour of data more accurately [26]. While the framework presented in this paper can be extended to cater for mixed Gaussian data, we assume that a single multivariate Gaussian distribution can represent the data.

### 3.2 Proposed DSSE Formulation

In order to estimate states, WLS algorithm [27] works based on minimization of the error between the measured data ($\boldsymbol{z}$) and a function ($\boldsymbol{h}(\boldsymbol{x})$) representing the relation between states ($\boldsymbol{x}$) and real measurements ($\mathbf{z} = \boldsymbol{h}(\boldsymbol{x})$) [28]. Usually, the states are bus voltage magnitudes and angles, while the measured values are active and reactive power injections and the reference bus voltage. WLS employs an iterative algorithm to estimate states leading to high computational cost. In contrast, our proposed method is a non-iterative algorithm using a direct approach to estimate the states of the network. In the proposed method, the real measured data is from measurement devices, whereas pseudo data are generated based on historical data. We use, injected currents ($\boldsymbol{i_{inj}}$) to estimate both bus voltages ($\boldsymbol{v_{inj}}$) and branch currents ($\boldsymbol{i_{branch}}$). Based on the direct power flow algorithm [29], the relationship between system states is given as:

$$\begin{cases} \boldsymbol{v_{inj}} = -\boldsymbol{DLF} \times \boldsymbol{i_{inj}} + \boldsymbol{v_{ref}} \\ \boldsymbol{i_{branch}} = \boldsymbol{BIBC} \times \boldsymbol{i_{inj}} \end{cases} \quad (4)$$

where $\boldsymbol{v_{ref}}$ is the voltage of the reference bus, $\boldsymbol{DLF}$ is the direct load flow matrix and $\boldsymbol{BIBC}$ is the bus-injection to the branch-current matrix. $\boldsymbol{DLF}$ and $\boldsymbol{BIBC}$ describe the network connection using line impedances. Following (4), the network states can directly be estimated as:

$$\begin{bmatrix} \boldsymbol{i_{inj}} \\ \boldsymbol{i_{branch}} \\ \boldsymbol{v_{inj}} \end{bmatrix} = \begin{bmatrix} \boldsymbol{I_{n-1}} & & \\ & \boldsymbol{BIBC} & \\ & & -\boldsymbol{DLF} \end{bmatrix} \begin{bmatrix} \boldsymbol{i_{inj}} \\ \boldsymbol{i_{inj}} \\ \boldsymbol{i_{inj}} \end{bmatrix} + \begin{bmatrix} \mathbf{0} \\ \mathbf{0} \\ \boldsymbol{v}_{\mathbf{ref}} \end{bmatrix} \quad (5)$$

where $\boldsymbol{I_{n-1}}$ is a unity square matrix of size $(n-1) \times (n-1)$, and $n$ is the number of buses.

A DSSE formulation is proposed to incorporate correlations between customer loads in different geographical locations and rolling time steps. In the proposed framework, $\boldsymbol{\Gamma}$ and $\boldsymbol{C}$ include spatial and temporal dependencies simultaneously. Before explaining the proposed method, let us introduce the correlation matrix ($\boldsymbol{CR}$) containing the relation between active powers ($\boldsymbol{CR_{PP}}$), active and reactive powers ($\boldsymbol{CR_{PQ}}$) and ($\boldsymbol{CR_{QP}}$), and reactive powers ($\boldsymbol{CR_{QQ}}$) for different locations and several time steps $(t, t-nt)$ as given in (6). There are four square matrices in (6) with $(n-1)\times nt$ size, where $nt$ represents the number of time steps considered.

$$\boldsymbol{CR} = \begin{bmatrix} \boldsymbol{CR_{P_tP_t}} & \cdots & \boldsymbol{CR_{P_tP_{t-nt}}} & \boldsymbol{CR_{P_tQ_t}} & \cdots & \boldsymbol{CR_{P_tQ_{t-nt}}} \\ \vdots & \ddots & \vdots & \vdots & \ddots & \vdots \\ \boldsymbol{CR_{P_{t-nt}P_t}} & \cdots & \boldsymbol{CR_{P_{t-nt}P_{t-nt}}} & \boldsymbol{CR_{P_{t-nt}Q_t}} & \cdots & \boldsymbol{CR_{P_{t-nt}Q_{t-nt}}} \\ \boldsymbol{CR_{Q_tP_t}} & \cdots & \boldsymbol{CR_{Q_tP_{t-nt}}} & \boldsymbol{CR_{Q_tQ_t}} & \cdots & \boldsymbol{CR_{Q_tQ_{t-nt}}} \\ \vdots & \ddots & \vdots & \vdots & \ddots & \vdots \\ \boldsymbol{CR_{Q_{t-nt}P_t}} & \cdots & \boldsymbol{CR_{Q_{t-nt}P_{t-nt}}} & \boldsymbol{CR_{Q_{t-nt}Q_t}} & \cdots & \boldsymbol{CR_{Q_{t-nt}Q_{t-nt}}} \end{bmatrix} \quad (6)$$

In order to provide a clearer and direct description of the proposed method, the framework is explained on a step-by-step base. The proposed DSSE algorithm is implemented through the following steps:

**Step 1:** Input the pseudo and real measurement data to calculate the spatial-temporal correlation matrix based on (1) and (6). The network schematics and line impedances are used to compute $\boldsymbol{BIBC}$ and $\boldsymbol{DLF}$ matrices.

**Step 2:** Calculate the mean and standard deviation (SD) of the real and imaginary parts of all injected currents. The mean values of the injected currents are calculated directly from the historical data using active and reactive power. It is assumed that the injected currents are calculated with respect to the voltage angle of the reference bus. Alternatively, the computed voltage angle obtained from running power flow using the pseudo measurement data, or the angle of the nearest measurement point can be used as the voltage angle of each bus [30]. The SD of the real and imaginary parts of the pseudo measurement data can be calculated based on the mean values and the measurement errors using (7) [31]:

$$\boldsymbol{\sigma^{Re}}_i + j\boldsymbol{\sigma^{Im}}_i = \left(\boldsymbol{\mu^{Re}}_i + j\boldsymbol{\mu^{Im}}_i\right) * \varepsilon/300 \quad (7)$$

where $\boldsymbol{\sigma^{Re}}$ and $\boldsymbol{\sigma^{Im}}$ are the real and imaginary parts of the SD, $\boldsymbol{\mu^{Re}}$ and $\boldsymbol{\mu^{Im}}$ are the real and imaginary parts of the mean values, $\varepsilon$ is measurement error, $i$ is the bus number and $j^2 = -1$.

**Step 3:** Calculate the covariance between the real and imaginary parts of the injected currents for each bus, and the covariance and pseudo covariance matrices based on the correlation matrix and SD using (8) [32]:

$$
\begin{aligned}
&\boldsymbol{COV_{Re,Re}} = \boldsymbol{\sigma^{Re}} * \boldsymbol{\sigma^{Re}} * \boldsymbol{CR_{PP}} \\
&\boldsymbol{COV_{Im,Im}} = \boldsymbol{\sigma^{Im}} * \boldsymbol{\sigma^{Im}} * \boldsymbol{CR_{QQ}} \\
&\boldsymbol{COV_{Re,Im}} = \boldsymbol{COV_{Im,Re}} = \boldsymbol{\sigma^{Re}} * \boldsymbol{\sigma^{Im}} * \boldsymbol{CR_{PQ}} \\
&\boldsymbol{\Gamma} = \boldsymbol{COV_{Re,Re}} + \boldsymbol{COV_{Im,Im}} + j(\boldsymbol{COV_{Im,Re}} - \boldsymbol{COV_{Re,Im}}) \\
&\boldsymbol{C} = \boldsymbol{COV_{Re,Re}} + \boldsymbol{COV_{Im,Im}} + j(\boldsymbol{COV_{Im,Re}} + \boldsymbol{COV_{Re,Im}})
\end{aligned}
\tag{8}
$$

where $\boldsymbol{COV_{Re,Re}}$ is the covariance between active powers, $\boldsymbol{COV_{Im,Im}}$ is the covariance between reactive powers, $\boldsymbol{COV_{Re,Im}}$ and $\boldsymbol{COV_{Im,Re}}$ are the covariance between active and reactive power.

A related point to consider is that a positive covariance between two RCs is referred to the higher correlation between the aggregated customer loads of these two RCs. Theoretically, two RCs with a positive covariance, are both likely to move in the same direction when responding to the changes in their customer loads behaviour. Hence, a higher correlation as a result of the load aggregation is employed in the CMCGD algorithm to update the pseudo data of the unmeasured RCs with the data of the monitored ones.

**Step 4:** In order to incorporate spatial-temporal correlation in the DSSE algorithm, the formulation in (9) is proposed to replace (5):

$$
\begin{bmatrix} \boldsymbol{i_{inj_{nt}}} \\ \boldsymbol{i_{branch_{nt}}} \\ \boldsymbol{v_{inj_{nt}}} \end{bmatrix} = \begin{bmatrix} \boldsymbol{I_{nt \times (n-1)}} & & \\ & \boldsymbol{BIBC_{nt}} & \\ & & \boldsymbol{-DLF_{nt}} \end{bmatrix} \begin{bmatrix} \boldsymbol{i_{inj_{nt}}} \\ \boldsymbol{i_{inj_{nt}}} \\ \boldsymbol{i_{inj_{nt}}} \end{bmatrix} + \begin{bmatrix} \boldsymbol{0} \\ \boldsymbol{0} \\ \boldsymbol{v_{\mathrm{ref}_{nt}}} \end{bmatrix} \tag{9}
$$

where $\boldsymbol{I}_{\boldsymbol{nt}\times(\boldsymbol{n}-\boldsymbol{1})}$ is an unity matrix and $\boldsymbol{v}_{\boldsymbol{ref}_{\boldsymbol{nt}}}$ is the measured voltage at the reference bus in each time step. $\boldsymbol{i}_{\boldsymbol{inj}_{\boldsymbol{nt}}}$, $\boldsymbol{i}_{\boldsymbol{branch}_{\boldsymbol{nt}}}$ and $\boldsymbol{v}_{\boldsymbol{inj}_{\boldsymbol{nt}}}$ are network states in each time step. $\boldsymbol{BIBC}_{\boldsymbol{nt}}$ and $\boldsymbol{DLF}_{\boldsymbol{nt}}$ are calculated using (10):

$$\boldsymbol{BIBC}_{\boldsymbol{nt}} = \begin{bmatrix} \boldsymbol{BIBC} & \cdots & \boldsymbol{0} \\ \boldsymbol{0} & \ddots & \boldsymbol{0} \\ \boldsymbol{0} & \cdots & \boldsymbol{BIBC} \end{bmatrix}, \boldsymbol{DLF}_{\boldsymbol{nt}} = \begin{bmatrix} \boldsymbol{DLF} & \cdots & \boldsymbol{0} \\ \boldsymbol{0} & \ddots & \boldsymbol{0} \\ \boldsymbol{0} & \cdots & \boldsymbol{DLF} \end{bmatrix} \tag{10}$$

where $\boldsymbol{BIBC}$ and $\boldsymbol{DLF}$ are repeated $n_t$ times as the elements of the diagonals of $\boldsymbol{BIBC}_{\boldsymbol{nt}}$ and $\boldsymbol{DLF}_{\boldsymbol{nt}}$, respectively.

**Step 5:** Denote $\boldsymbol{\mu}_{\boldsymbol{t}}$ and $\boldsymbol{\mu}_{\boldsymbol{vreft}}$ as the mean injected currents and voltage at reference bus, in time step $t$. The mean injected and branch currents and bus voltages ($\boldsymbol{\mu}_{\boldsymbol{IBVt}}$) are given as:

$$\boldsymbol{\mu}_{\boldsymbol{IBV}_{\boldsymbol{t}}} = \begin{bmatrix} \boldsymbol{\mu}_{\boldsymbol{t}} \\ \boldsymbol{BIBC}_{\boldsymbol{nt}} * \boldsymbol{\mu}_{\boldsymbol{t}} \\ -\boldsymbol{DLF}_{\boldsymbol{nt}} * \boldsymbol{\mu}_{\boldsymbol{t}} + \boldsymbol{\mu}_{\boldsymbol{vreft}} \end{bmatrix} \tag{11}$$

where *IBV* stands for injected/branch currents/voltage of the buses.

**Step 6:** Based on the matrices $\boldsymbol{BIBC}_{\boldsymbol{nt}}$ and $\boldsymbol{DLF}_{\boldsymbol{nt}}$, the covariance and pseudo covariance matrices ($\boldsymbol{\Gamma}_{\boldsymbol{IBV}}$ and $\boldsymbol{C}_{\boldsymbol{IBV}}$) characterizing the statistical behavior of all the states of the network, and for all time steps are given in (12).

$$\begin{gathered} \boldsymbol{\Gamma}_{\boldsymbol{IBV}} = \begin{bmatrix} \boldsymbol{\Gamma} & \boldsymbol{\Gamma} * \boldsymbol{BIBC}_{\boldsymbol{nt}}^{\boldsymbol{H}} & -\boldsymbol{\Gamma} * \boldsymbol{DLF}_{\boldsymbol{nt}}^{\boldsymbol{H}} \\ \boldsymbol{BIBC}_{\boldsymbol{nt}} * \boldsymbol{\Gamma} & \boldsymbol{BIBC}_{\boldsymbol{nt}} * \boldsymbol{\Gamma} * \boldsymbol{BIBC}_{\boldsymbol{nt}}^{\boldsymbol{H}} & -\boldsymbol{BIBC}_{\boldsymbol{nt}} * \boldsymbol{\Gamma} * \boldsymbol{DLF}_{\boldsymbol{nt}}^{\boldsymbol{H}} \\ \boldsymbol{DLF}_{\boldsymbol{nt}} * \boldsymbol{\Gamma} & -\boldsymbol{DLF}_{\boldsymbol{nt}} * \boldsymbol{\Gamma} * \boldsymbol{BIBC}_{\boldsymbol{nt}}^{\boldsymbol{H}} & \boldsymbol{DLF}_{\boldsymbol{nt}} * \boldsymbol{\Gamma} * \boldsymbol{DLF}_{\boldsymbol{nt}}^{\boldsymbol{H}} + \boldsymbol{\Gamma}_{\boldsymbol{vref}} \end{bmatrix} \\ \boldsymbol{C}_{\boldsymbol{IBV}} = \begin{bmatrix} \boldsymbol{C} & \boldsymbol{C} * \boldsymbol{BIBC}_{\boldsymbol{N}}^{\boldsymbol{T}} & -\boldsymbol{C} * \boldsymbol{DLF}_{\boldsymbol{nt}}^{\boldsymbol{T}} \\ \boldsymbol{BIBC}_{\boldsymbol{nt}} * \boldsymbol{C} & \boldsymbol{BIBC}_{\boldsymbol{nt}} * \boldsymbol{C} * \boldsymbol{BIBC}_{\boldsymbol{nt}}^{\boldsymbol{T}} & -\boldsymbol{BIBC}_{\boldsymbol{nt}} * \boldsymbol{C} * \boldsymbol{DLF}_{\boldsymbol{nt}}^{\boldsymbol{T}} \\ \boldsymbol{DLF}_{\boldsymbol{nt}} * \boldsymbol{C} & -\boldsymbol{DLF}_{\boldsymbol{nt}} * \boldsymbol{C} * \boldsymbol{BIBC}_{\boldsymbol{nt}}^{\boldsymbol{T}} & \boldsymbol{DLF}_{\boldsymbol{nt}} * \boldsymbol{C} * \boldsymbol{DLF}_{\boldsymbol{nt}}^{\boldsymbol{T}} + \boldsymbol{C}_{\boldsymbol{vref}} \end{bmatrix} \end{gathered} \tag{12}$$

**Step 7:** By considering temporal correlation, (3) is rewritten as:

$$\begin{gathered} \boldsymbol{\mu}_{\boldsymbol{2}}{}^{\boldsymbol{updated}} = \boldsymbol{\mu}_{\boldsymbol{2}} + \sum_{k=1}^{nt} [\boldsymbol{A}\left(\boldsymbol{\mu}_{\boldsymbol{1}}^{\boldsymbol{M}}(k,) - \boldsymbol{\mu}_{\boldsymbol{1}}(k,)\right) + \boldsymbol{B}\left(\boldsymbol{\mu}_{\boldsymbol{1}}^{\boldsymbol{M}}(k,) - \boldsymbol{\mu}_{\boldsymbol{1}}(k,)\right)^{*}] \\ \boldsymbol{\Gamma}_{\boldsymbol{22nt}}{}^{\boldsymbol{updated}} = \boldsymbol{\Gamma}_{\boldsymbol{22nt}} - \boldsymbol{A}\boldsymbol{\Gamma}_{\boldsymbol{12nt}}^{\boldsymbol{H}} - \boldsymbol{B}\boldsymbol{\Gamma}_{\boldsymbol{12nt}}^{\boldsymbol{H}} \\ \boldsymbol{C}_{\boldsymbol{22nt}}{}^{\boldsymbol{updated}} = \boldsymbol{C}_{\boldsymbol{22nt}} - \boldsymbol{A}\boldsymbol{C}_{\boldsymbol{12nt}}^{\boldsymbol{T}} - \boldsymbol{B}\boldsymbol{C}_{\boldsymbol{12nt}}^{\boldsymbol{T}} \end{gathered} \tag{13}$$

where $(,)$ represents an element of a vector. The results later given in Section 5 highly support the incorporation of temporal correlation in reducing estimation error as proposed in (13). In this equation, the measured injected and branch currents and node voltages can be considered to update pseudo data. In the distribution networks with a low number of measurement devices, the proposed algorithm employs the previous time steps for accurately updating the pseudo data for the DSSE algorithm, as shown in Figure 5.

**Step 8:** After updating the mean matrix of pseudo injected currents using (13), calculate all states ($\boldsymbol{\mu_{IBV_t}}^{\boldsymbol{updated}}$) of the distribution network using (14), while instead of $\boldsymbol{\mu_t}$, the updated mean values ($\boldsymbol{\mu_t}^{\boldsymbol{updated}}$) are used. Furthermore, by updating the covariance ($\boldsymbol{\Gamma_{IBV}}^{\boldsymbol{updated}}$) and pseudo covariance ($\boldsymbol{C_{IBV}}^{\boldsymbol{updated}}$) matrices using (13), the real ($\boldsymbol{COV_{Re,Re}}^{\boldsymbol{updated}}$) and imaginary ($\boldsymbol{COV_{Im,Im}}^{\boldsymbol{updated}}$) parts of the covariance between states can be calculated as (15) [33]:

$$\boldsymbol{\mu_{IBV_t}}^{\boldsymbol{updated}} = \begin{bmatrix} \boldsymbol{\mu_t}^{\boldsymbol{updated}} \\ \boldsymbol{BIBC_{nt}} * \boldsymbol{\mu_t}^{\boldsymbol{updated}} \\ -\boldsymbol{DLF_{nt}} * \boldsymbol{\mu_t}^{\boldsymbol{updated}} + \boldsymbol{\mu_{vreft}} \end{bmatrix} \tag{14}$$

$$\begin{cases} \boldsymbol{COV_{Re,Re}}^{\boldsymbol{updated}} = {}^{1}/_{2}\, Real\left(\boldsymbol{\Gamma_{IBV}}^{\boldsymbol{updated}} + \boldsymbol{C_{IBV}}^{\boldsymbol{updated}}\right) \\ \boldsymbol{COV_{Im,Im}}^{\boldsymbol{updated}} = {}^{1}/_{2}\, Real\left(\boldsymbol{\Gamma_{IBV}}^{\boldsymbol{updated}} - \boldsymbol{C_{IBV}}^{\boldsymbol{updated}}\right) \\ \boldsymbol{COV_{Re,Im}}^{\boldsymbol{updated}} = {}^{1}/_{2}\, Imaginary\left(\boldsymbol{\Gamma_{IBV}}^{\boldsymbol{updated}} + \boldsymbol{C_{IBV}}^{\boldsymbol{updated}}\right) \\ \boldsymbol{COV_{Im,Re}}^{\boldsymbol{updated}} = {}^{1}/_{2}\, Imaginary\left(\boldsymbol{\Gamma_{IBV}}^{\boldsymbol{updated}} - \boldsymbol{C_{IBV}}^{\boldsymbol{updated}}\right) \end{cases} \tag{15}$$

**Step 9:** Finally, based on the results in the previous step, the variance ($\boldsymbol{VAR}$) of the magnitude and angle of the states can be written as in (16).

$$\begin{cases} \boldsymbol{VAR}(|\boldsymbol{X}|) = \boldsymbol{\mu}_{\boldsymbol{X^{RI}}}^{\boldsymbol{T}} * \boldsymbol{COV_{X^{RI},X^{RI}}} * \boldsymbol{\mu_{X^{RI}}} / \|\boldsymbol{X^{RI}}\|^2 \\ \boldsymbol{VAR}(\angle \boldsymbol{X}) = \boldsymbol{\mu}_{\boldsymbol{X^{IR}}}^{\boldsymbol{T}} * \boldsymbol{COV_{X^{RI},X^{RI}}} * \boldsymbol{\mu_{X^{IR}}} / \|\boldsymbol{X^{RI}}\|^4 \end{cases} \tag{16}$$

where $\boldsymbol{X} = \boldsymbol{X^{Re}} + j\boldsymbol{X^{Im}}$, $\boldsymbol{X^{RI}} = [\boldsymbol{X^{Re}}\ \boldsymbol{X^{Im}}]^T$, $\boldsymbol{X^{IR}} = [-\boldsymbol{X^{Im}}\ \boldsymbol{X^{Re}}]^T$, $\|\boldsymbol{X^{RI}}\|$ is $l_2$ norm of $\boldsymbol{X^{RI}}$, and

$$\boldsymbol{COV_{X^{RI},X^{RI}}} = \begin{bmatrix} \boldsymbol{COV_{Re,Re}}^{\boldsymbol{updated}} & \boldsymbol{COV_{Re,Im}}^{\boldsymbol{updated}} \\ \boldsymbol{COV_{Im,Re}}^{\boldsymbol{updated}} & \boldsymbol{COV_{Im,Im}}^{\boldsymbol{updated}} \end{bmatrix}.$$

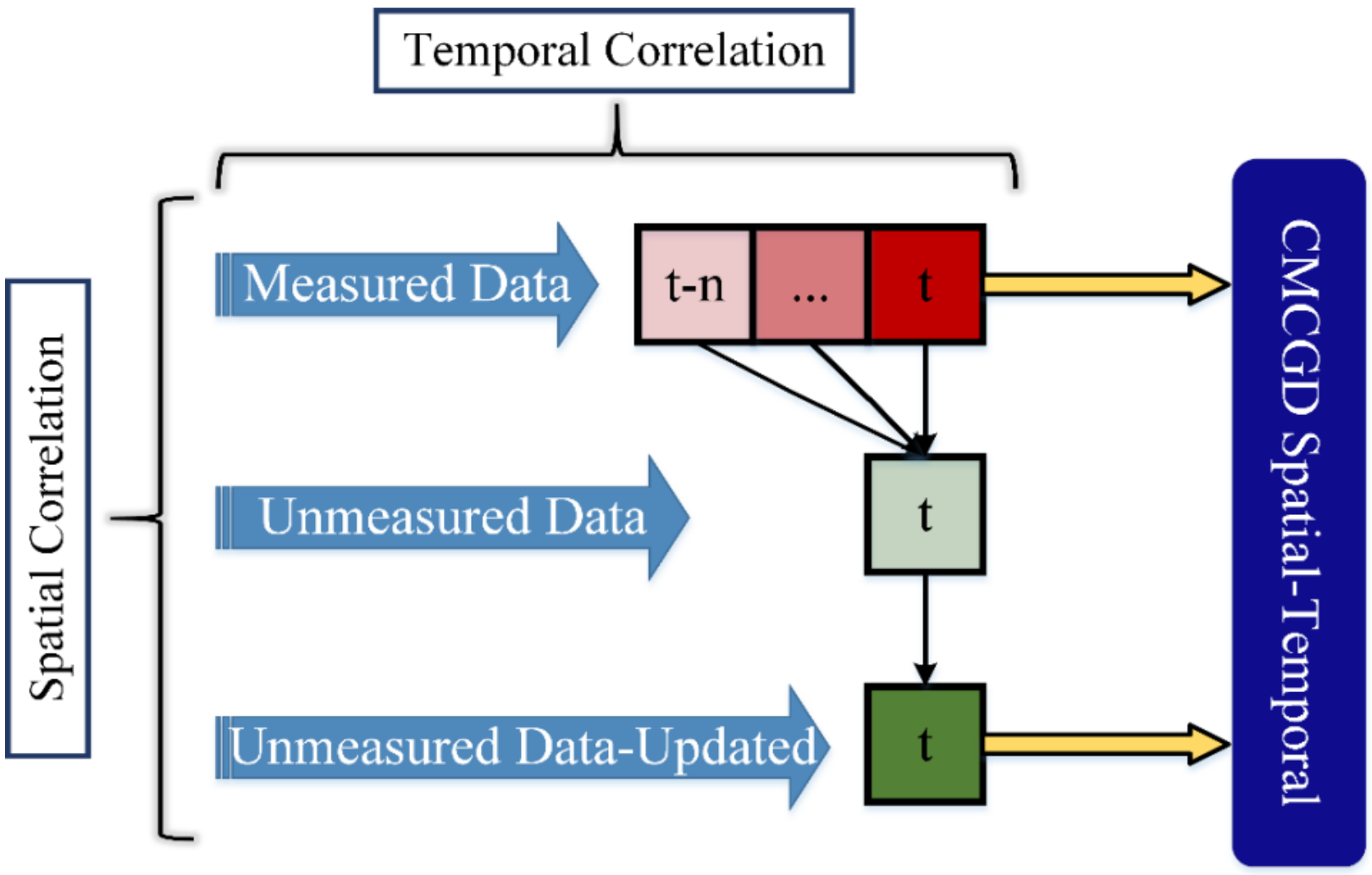

Figure. 5. Spatial-temporal correlation in CMCGD.

In order to decrease the computational complexity and the simulation time for the real-time applications, the first four steps can be computed off-line, considered as initial steps. The flowchart of the proposed method with separated off-line and real-time steps is presented in Figure 6. All off-line steps are calculated once before employing the proposed DSSE algorithm, while the real-time steps are updated in each time step.

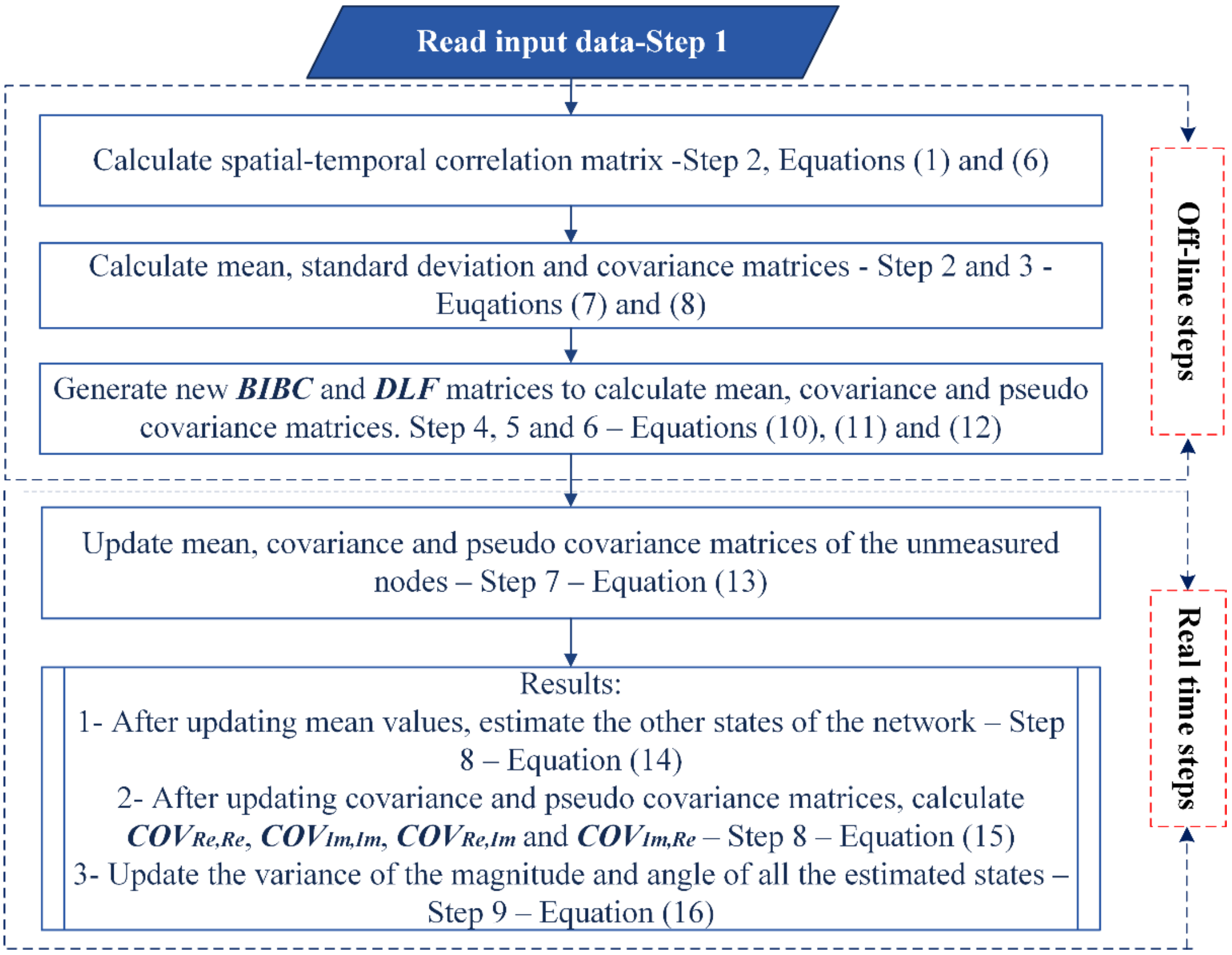

Figure. 6. The flowchart of the spatial-temporal CMCGD state estimator.

In the last part of this section, it is worth to note that increasing the number of time steps increases the computational time (especially for large distribution networks). Furthermore, as we go further back in time, the temporal correlation will decrease, and only contributes to further unnecessary computation time. Hence, a sensitivity analysis is required to find the most efficient number of considered time steps to capture the impact of temporal correlation. Based on our studies applied to two separate LV distribution networks, three time steps one minute each found to be the most optimal number.

## 4. Description of Data

In this study, smart meters used to collect data from one hundred houses with 15%-25% PV penetration rate in Newmarket suburb. Seven days' worth of data with one-minute resolution is used in this paper. Also, the data with one minute resolution from Pecan Street, Texas, USA [23] is considered in section 2.

It is worth to note that this complex DSSE algorithm requires the measured data with magnitude and angle. However, in a distribution network with traditional magnitude measurement devices, the calculated angle based on the pseudo data can be considered for the complex state estimation.

The SDs of the states are calculated based on the measurement errors, assuming 3% and 50% for the real and pseudo measured data, respectively.

It should be noted that the correlation matrix in this study should be positive definite. Therefore, in the cases of real data analysis, Rstudio is employed to find the nearest positive definite correlation matrix [34].

## 5. Simulation Results

In this section, four case studies are considered.  In the first case, a radial balanced medium voltage (MV) with six-bus distribution network is considered [29] and the performance of the proposed method is evaluated under several operating conditions. The second case study is a larger network; the balanced 123 node IEEE test MV feeder [35]. A 615 node test feeder with balanced customer loads is also considered as the third case study. Finally, in the fourth case study, the efficiency of the proposed algorithm is verified on a real Australian residential LV distribution network with unbalanced loads. In the last case study, the impact of increased R/X ratio is studied to measure the accuracy of the proposed method in the distribution networks with high R/X ratio cables. MATLAB on Intel Core i70-4600 with clock speed 2.7 GHz and 16 GB RAM is used for the simulations.

In order to examine the effect of the temporal correlation in the proposed CMCGD spatial-temporal (CST) method, the results of the estimated states are compared with the CMCGD spatial (CS) model and WLS algorithm. For a compatible comparison, the same real and pseudo measured data are used to implement in all three methods. Also, spatial correlation is considered to form the covariance matrix in the WLS algorithm. Rival approaches are compared based on the average magnitude voltage error (AMVE) and average angle voltage error (AAVE) [36], computational time, number of iterations and SD of the estimated states as the quality index [26]. AMVE and AAVE in the case studies of 1, 2 and 3 are the average error of bus voltages at single time steps, while they represent the average estimated voltage error at each bus for one full day in the case study 4. AMVE and AAVE criterions are more critical compared with the other three factors, because, in control and protection studies and applications, the accurate bus voltages are vital for the correct decision in operational procedures. Computational time and iteration are also playing an important role in real-time control and protection applications. Finally, the quality of the estimators has an inverse relationship with its states variance as given in (17):

$$Quality = ln(\frac{1}{tr(Cov_{X^{RI},X^{RI}})}) \tag{17}$$

where $tr(.)$ represents matrix trace.

### 5.1 Case Study 1: 6-bus distribution system

Figure 7 shows a 6-bus radial MV distribution network. The pseudo measurement of the injected current in each bus is provided in Table 1.

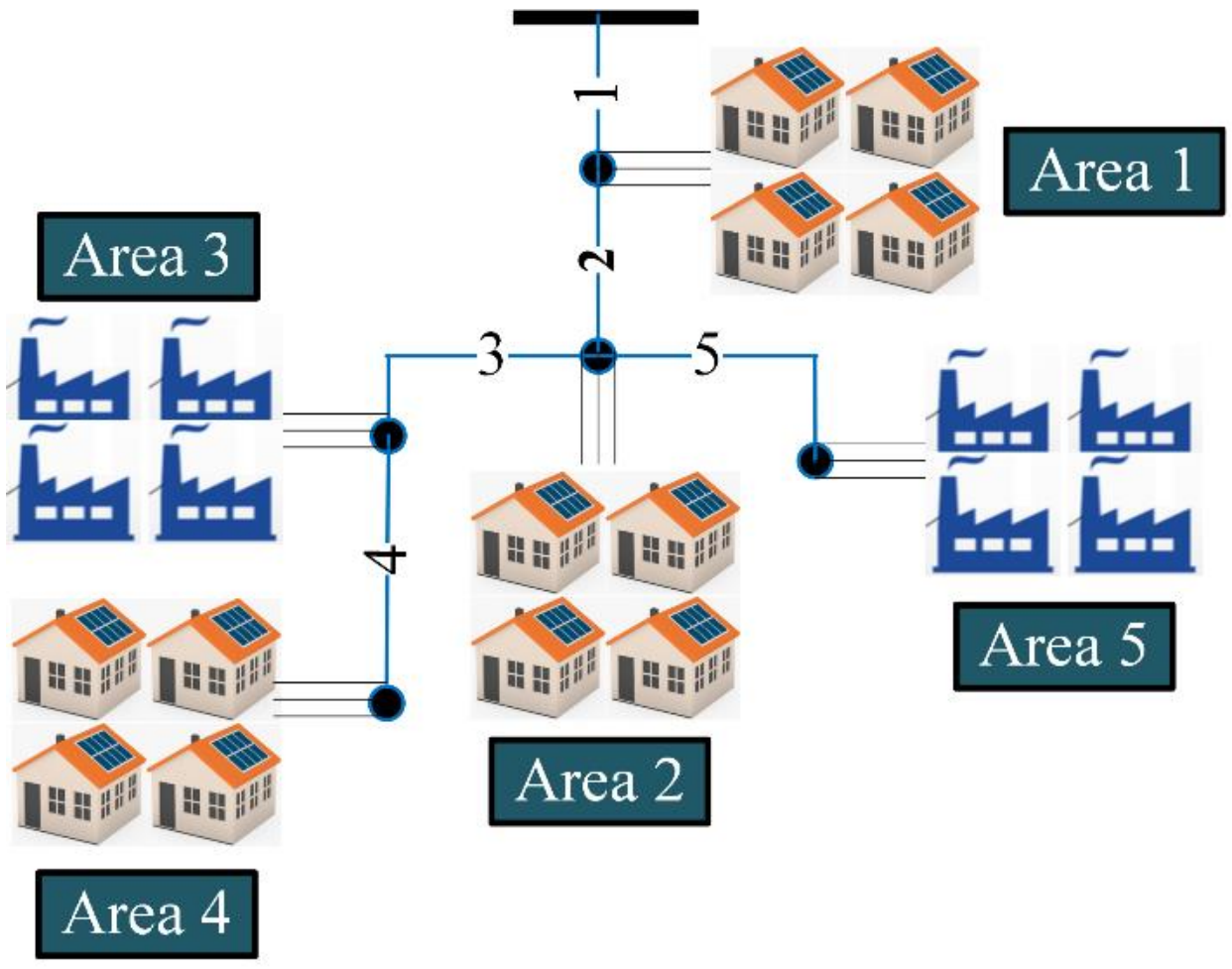

Figure 7. A six-bus distribution network.

Table 1. Pseudo measured injected current at 100% loading condition

| Area number | Area 1 | Area 2 | Area 3 | Area 4 | Area 5 |
|---|---|---|---|---|---|
| **Injected current(A)** | 18.9-6.7i | 13.4-12.7i | 14.8-10.9i | 16.8-14.9i | 17.7-14.9i |

In order to examine the impact of the correlation between the variables, two types of load batches are considered. Areas 1, 2 and 4 represent residential loads batches, while Areas 3 and 5 involve a combination of several industrial loads. In distribution networks, same load types usually show high cross-correlation. The correlation matrix in case 1 is visualized in Figure 8, where black to white colours show the highest to lowest correlations. In this case, three scenarios are considered, as described in the followings.

### 5.1.1 First Scenario

Only, Area 1 has an injection current measurement, and for the other buses, the injected current data at 100% loading is considered as pseudo measurements. In this scenario, as shown in the daily load profile in Figure 9, the loading of the system decreases in three steps by; 20%, 40%, and 60%, while only the data of Area 1 is updated in DSSE. The comparative results of the three methods are given in Table 2. As given in Table 2, WLS algorithm updates only the loading of Area 1 and it fails to update the customer loads at the rest of the Areas as well as the error of the pseudo measured data. This leads to 6.76 quality index for WLS versus 7.41 and 7.42 quality indices for CS and CST.

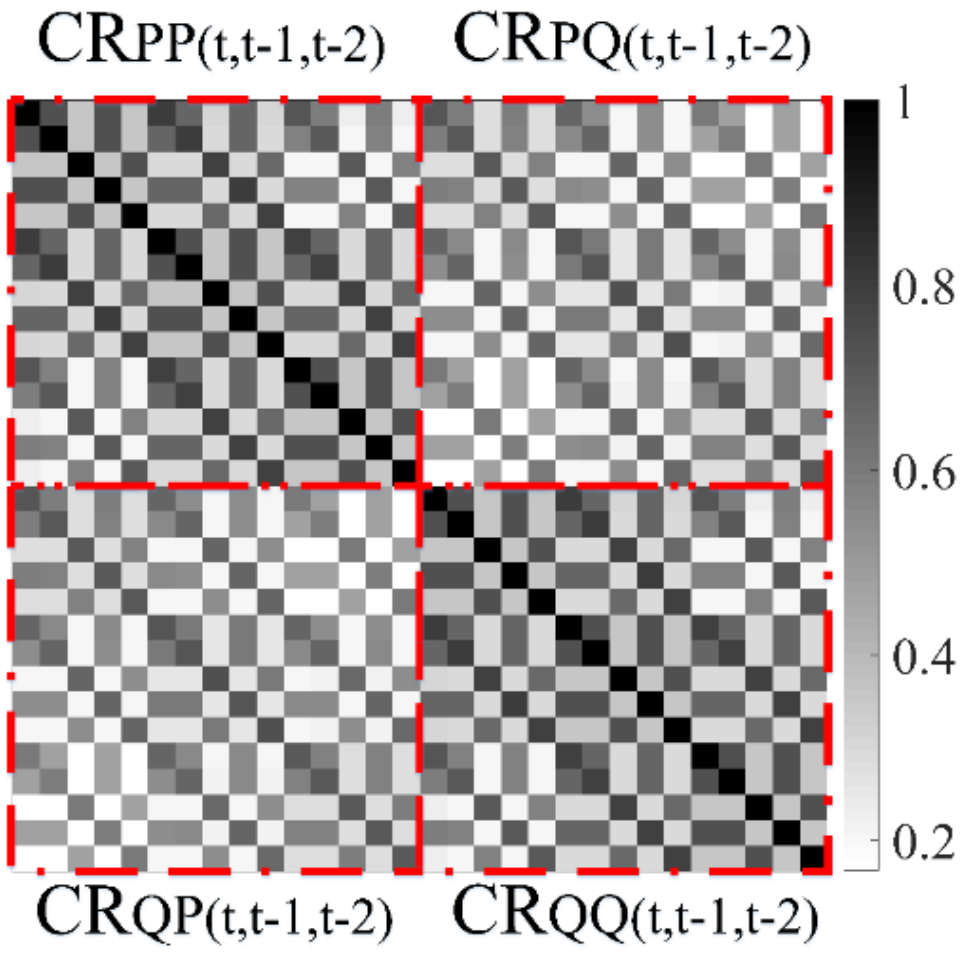

Figure 8. Visualization of the correlation matrix for case study 1.

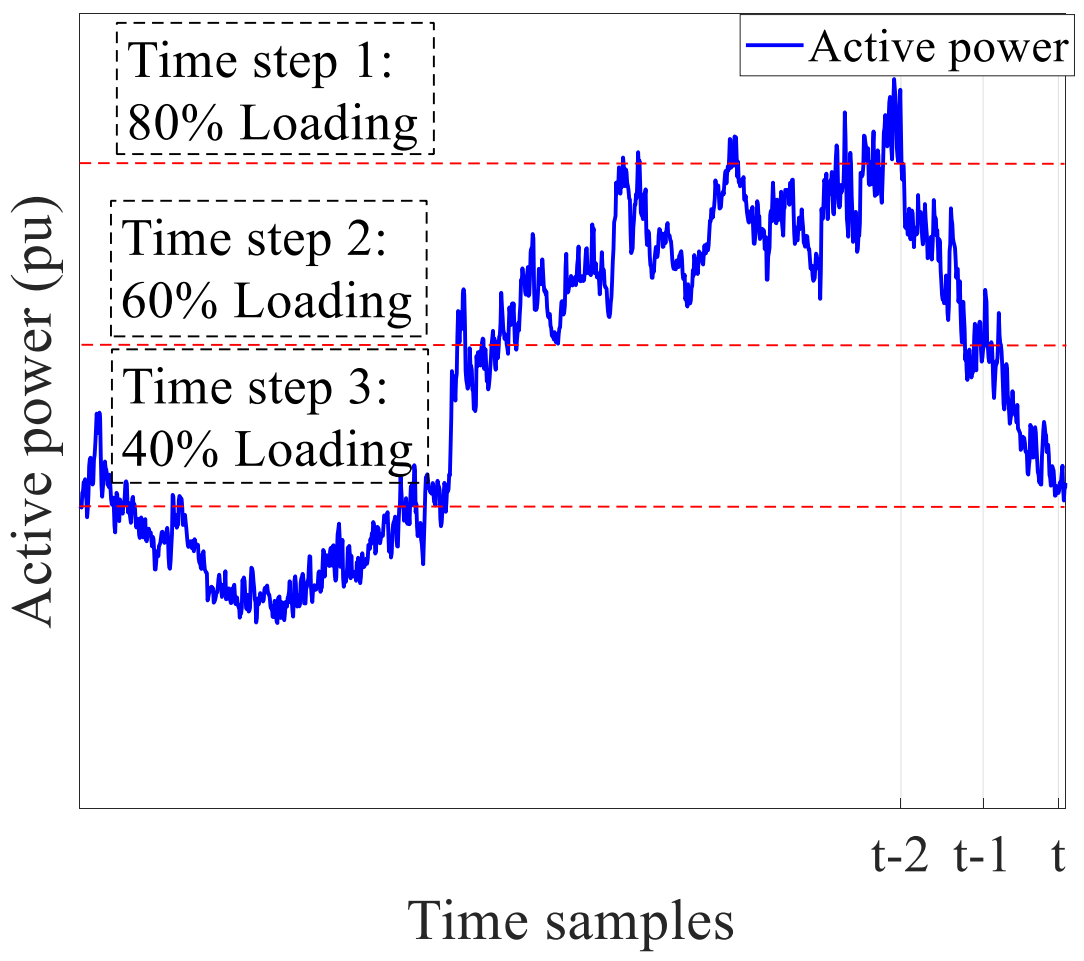

Figure 9. Real-time daily load profile for case study 1.

Table 2. Comparative results from Case study 1, scenario 1 – for three time steps and three loading levels

| | Time step 1: 80% Loading | | | Time step 2: 60% Loading | | | Time step 3: 40% Loading | | |
|---|---|---|---|---|---|---|---|---|---|
| | **WLS** | **CS** | **CST** | **WLS** | **CS** | **CST** | **WLS** | **CS** | **CST** |
| **AMVE (%)** | 3.9 | 0.3 | 0.3 | 4.8 | 0.9 | 0.7 | 6.1 | 1.4 | 1.1 |
| **AAVE (deg)** | 0.06 | 0.01 | 0.01 | 0.05 | 0.01 | 0.009 | 0.05 | 0.01 | 0.007 |
| **Quality** | 6.76 | 7.41 | 7.41 | 6.76 | 7.41 | 7.42 | 6.76 | 7.41 | 7.42 |
| **Time(s)** | 0.03 | 0.01 | 0.01 | 0.03 | 0.01 | 0.01 | 0.03 | 0.01 | 0.01 |
| **Iterations** | 5 | 1 | 1 | 5 | 1 | 1 | 5 | 1 | 1 |

The iterative nature of a method potentially increases the computational time. This argument is supported in Table 2 where WLS needs 0.03s simulation time, while CS and CST require only 0.01s for computations. In the first time step, the load level is 80% and only one new measured data is imported into the algorithms. Hence, the temporal correlation cannot be considered here and the output of CS and CST are the same. The magnitude and angle of AMVE and AAVE in WLS are 3.9% and 0.06 degree while they are 0.3% and 0.01 degree for CS and CST, in the first time step. The reason for this difference is that in the WLS algorithm, the statistical parameters of pseudo measurements cannot be updated based on the measured input. For the next time step, the system operates at 60% loading, in which the loading of Area 1 is updated in all algorithms. In CST, the measured value at step one is kept the same for step 2. By reducing the loading of the system to 60%, the difference between available pseudo data and the real measurement is increased, which causes a 0.9% rise in the magnitude of AMVE in WLS. As given in Table 2, the magnitude of AMVE in CST is 0.7%, which is lower than that of CS. This improvement is achieved by making use of temporal correlations between the two successive time steps in CST according to the proposed formulation in (13). In the last time step, the 6-bus system operates at 40% loading. Again, for CST, the measured data in time step 3 is kept the same as that in the previous time steps. As it is expected, the average magnitude error in WLS is increased to 6.1%. The same error index is 1.4% and 1.1% for CS and CST, respectively. This supports the idea of integrating temporal and spatial correlations into the design of state estimation algorithms.

#### 5.1.2 Second Scenario

The loading levels are 80% and 40% in the first two steps and 60% in the last step. Hence, the system experiences first a decrease followed by an increase in the load level. The empirical results are given in Table 3. From Table 3, one concludes that CST performs better than CS and WLS. As seen in Table 3, CST results in 1.2% and 0.5% average voltage magnitude errors at 40% and 60% loading, respectively, while the same error indices are 1.4% and 0.9% in CS algorithm and 6.1% and 4.8% in WLS algorithm. It is interesting to note the impact of including more time steps. For instance, in the first scenario, the AMVE in the CST algorithm at 60% loading (the second time step) is 0.7%, while in the second scenario for the same level of loading, the error decreases to 0.5% at the third time step.

#### 5.1.3 Third Scenario

As noted before, two different load types namely residential and industrial are considered in the first case study. These load types are not highly correlated. The industrial loads come to the picture in the third scenario where another injected current measurement device is considered in the industrial load group on Area 3. The aim of designing this scenario is to show that adding spatially distributed measurement points leads to a significant reduction in the estimation error. Table 4 summarizes the results for three time steps. As shown in Figure 10, by adding another measurement point to WLS algorithm, due to neglecting correlations between load groups, the magnitude of AMVE is only slightly decreased. On the other hand, adding one more measurement device in one of the industrial load groups significantly improve the performance of CS and CST.

Table 3. Comparison results for scenario 2 – for three time steps with a decrease and an Increase in loading

| | **Time step 1: 80% Loading** | | | **Time step 2: 40% Loading** | | | **Time step 3: 60% Loading** | | |
|---|---|---|---|---|---|---|---|---|---|
| | **WLS** | **CS** | **CST** | **WLS** | **CS** | **CST** | **WLS** | **CS** | **CST** |
| **AMVE (%)** | 3.9 | 0.3 | 0.3 | 6.1 | 1.4 | 1.2 | 4.8 | 0.9 | 0.5 |
| **AAVE (deg)** | 0.06 | 0.01 | 0.01 | 0.05 | 0.01 | 0.008 | 0.05 | 0.01 | 0.007 |
| **Quality** | 6.76 | 7.41 | 7.41 | 6.76 | 7.41 | 7.42 | 6.76 | 7.41 | 7.42 |
| **Time(s)** | 0.03 | 0.01 | 0.01 | 0.03 | 0.01 | 0.01 | 0.03 | 0.01 | 0.01 |
| **Iterations** | 5 | 1 | 1 | 5 | 1 | 1 | 5 | 1 | 1 |

Table 4. Comparison results for scenario 3 with two measurement points and three Time steps

| | **Time step 1: 80% Loading** | | | **Time step 2: 60% Loading** | | | **Time step 3: 40% Loading** | | |
|---|---|---|---|---|---|---|---|---|---|
| | WLS | CS | CST | WLS | CS | CST | WLS | CS | CST |
| **AMVE (%)** | 3.6 | 0.1 | 0.1 | 4.1 | 0.4 | 0.3 | 5.4 | 0.7 | 0.6 |
| **AAVE (deg)** | 0.05 | 0.009 | 0.009 | 0.04 | 0.008 | 0.007 | 0.04 | 0.008 | 0.006 |
| **Quality** | 6.78 | 7.42 | 7.42 | 6.81 | 7.42 | 7.42 | 6.81 | 7.42 | 7.42 |
| **Time(s)** | 0.03 | 0.01 | 0.01 | 0.03 | 0.01 | 0.01 | 0.03 | 0.01 | 0.01 |
| **Iterations** | 5 | 1 | 1 | 5 | 1 | 1 | 5 | 1 | 1 |

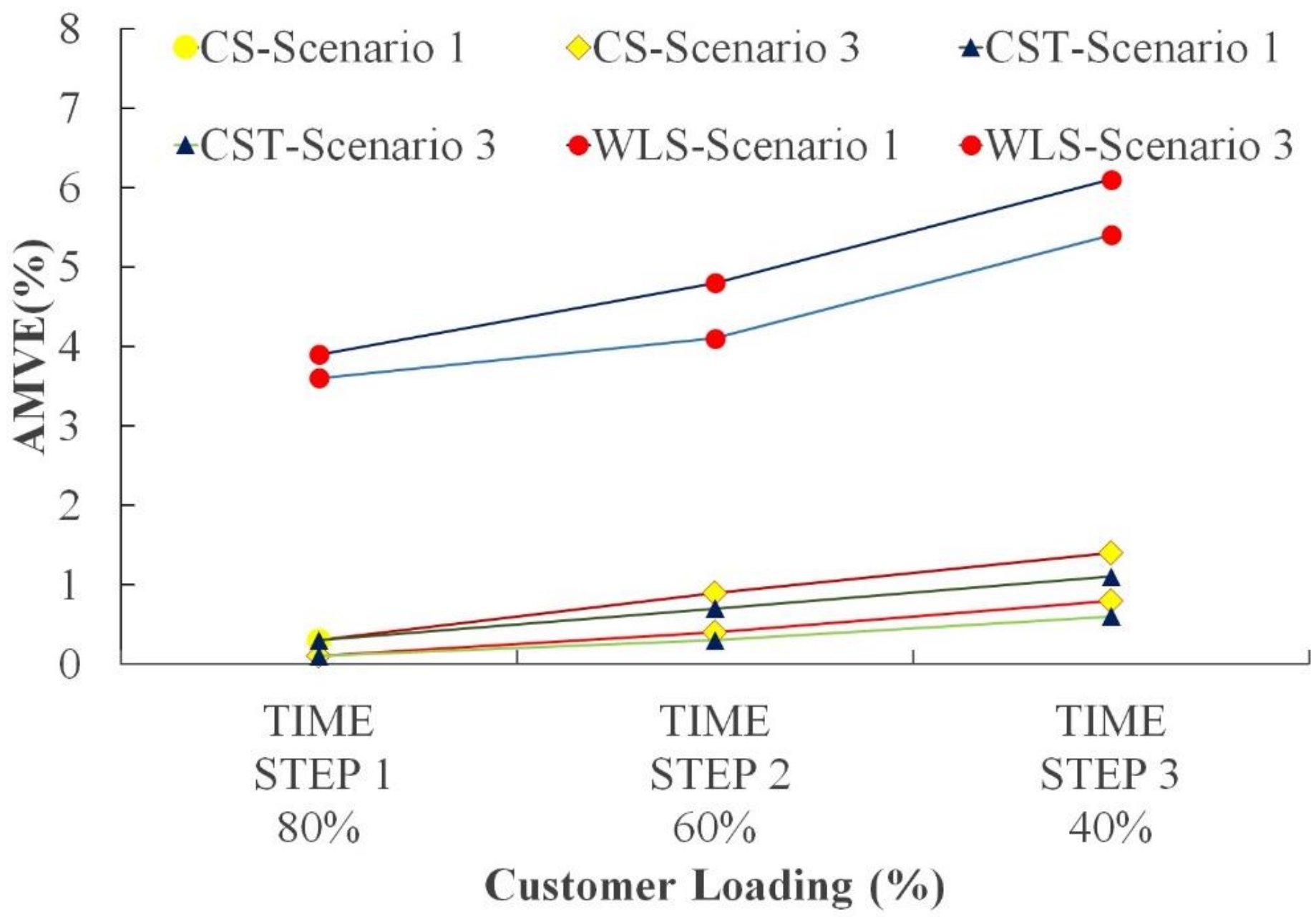

Figure 10. Average voltage errors for scenario 1 and 3.

In CS, the magnitudes of AMVE in the first scenario are 0.3%, 0.9%, and 1.4%, respectively, while by importing a new measured data on Area 3, errors are decreased to 0.1%, 0.4% and 0.8% for three time steps. The improvements are more significant in CST case where the errors of voltage estimation are decreased to 0.1%, 0.3% and 0.6% for the three successive time steps.

**5.2 Case Study 2: IEEE 123 Node Test Feeder**

In order to verify the performance of the proposed method in a larger distribution network, the IEEE test system with 123 nodes as shown in Figure 11 is considered. The load data is assumed as pseudo measured data at 100% loading [37]. The blue area shows the group of residential loads, while the green areas represent three industrial regions. In this case study, sufficient historical data is not available to calculate the correlation matrix. However, because the load levels and types are similar to those in case study 1, we assume the same correlation coefficients are present in these two case studies. Six injected current measurements are assumed, three of them are installed in industrial regions, on buses 28, 77 and 109. For residential loads, three measurement points are considered on buses 2, 48 and 69. Three operating load points with 80%, 60% and 40% in three time steps are studied in this case. The comparative results are summarized in Table 5.

From Table 5, one concludes that WLS imposes much higher computational cost comparing to CS and CST. Simulation time required for WLS is 0.12s while it is only 0.04s for CS and CST. As expected, by increasing the number of buses, the quality of the state estimation decreases to 2.33, 4.48 and 4.48 for WLS, CS, and CST, respectively. Furthermore, the AMVE and AAVE for WLS are about 2% and 0.018 degree while they are 0.09% and 0.001 degree for CS and CST.

Table 5. Comparison results for case study 2, with decreasing load

| | **Time step 1: 80% Loading** | | | **Time step 2: 60% Loading** | | | **Time step 3: 40% Loading** | | |
|---|---|---|---|---|---|---|---|---|---|
| | WLS | CS | CST | WLS | CS | CST | WLS | CS | CST |
| **AMVE (%)** | 2.05 | 0.09 | 0.09 | 2.45 | 0.32 | 0.3 | 2.84 | 0.52 | 0.42 |
| **AAVE (deg)** | 0.018 | 0.001 | 0.001 | 0.024 | 0.003 | 0.002 | 0.028 | 0.005 | 0.004 |
| **Quality** | 2.33 | 4.48 | 4.48 | 2.42 | 4.48 | 4.48 | 2.42 | 4.48 | 4.49 |
| **Time(s)** | 0.12 | 0.04 | 0.04 | 0.12 | 0.04 | 0.05 | 0.12 | 0.04 | 0.07 |
| **Iterations** | 5 | 1 | 1 | 5 | 1 | 1 | 5 | 1 | 1 |

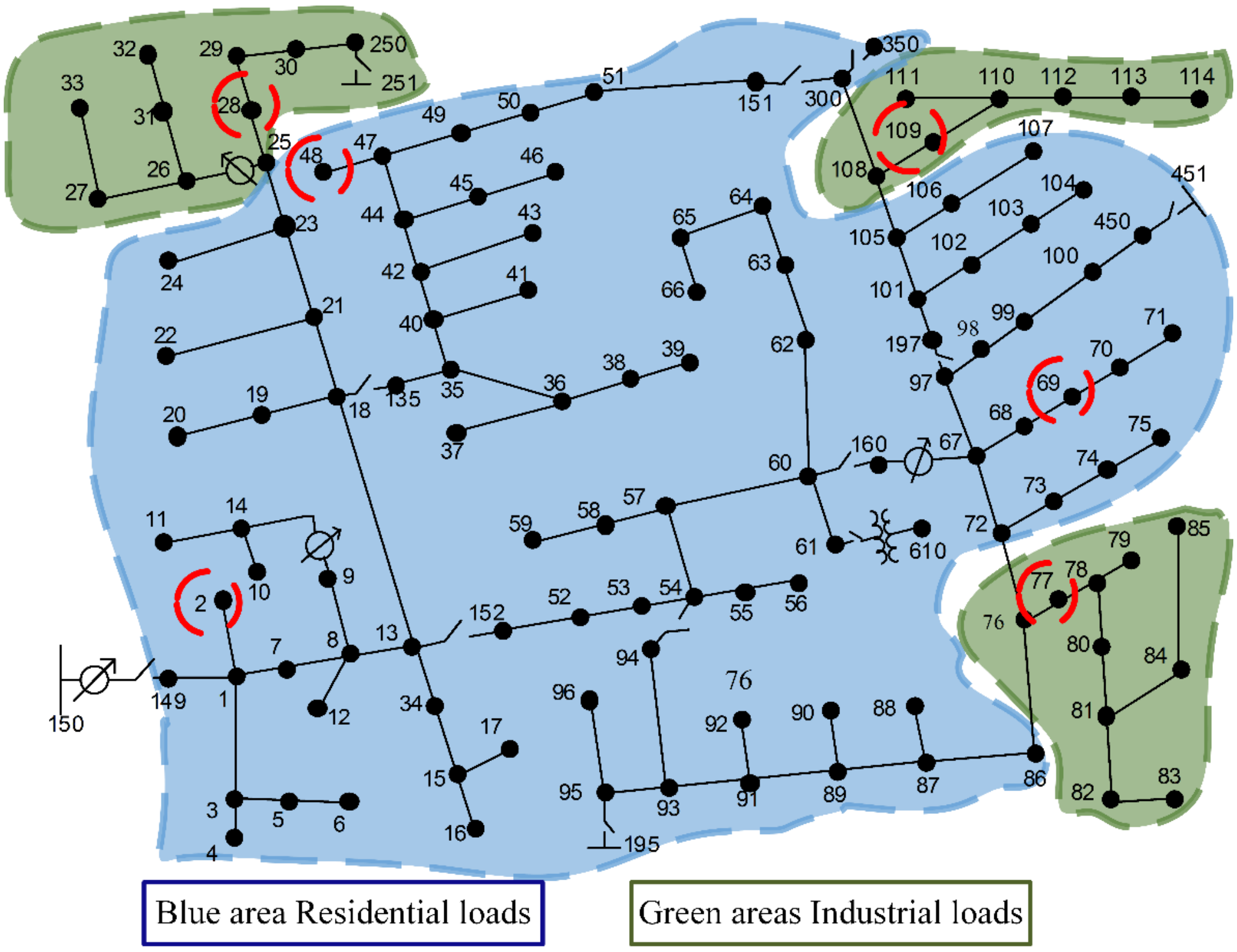

Figure 11. IEEE 123 node test feeder [37].

In the second and third time steps, the loading of the distribution network is 60% and 40%, respectively. The superiority of CST over CS in terms of the ability to reduce estimation error is implied in the second and third time steps. Although in these time steps, the magnitude and angle errors are increased for all algorithms, the CST shows the lowest error rates. This improvement is the result of incorporating both spatial and temporal correlations in the measured data. The simulation time in CST increases due to the fact that the amount of imported data into the algorithm is three times larger than that for the CS algorithm. The simulation time difference between these two methods is around 0.03s which is not significant and manageable for the application of state estimation to large distribution networks.

### 5.3 Case Study 3: 615 Node Test Feeder

In this section, the accuracy and the computational burden of the proposed method are evaluated in a large 615 node distribution network. It is worthwhile to point that the 615 node test feeder is constructed by five of the IEEE 123 node test feeders in parallel. Each of the 123 node network has the group of residential and industrial loads with 6 measurement devices. Hence, the 615 node distribution network operates with 30 measurement devices. The correlation coefficients in this case study are similar to the case study 2, and the 100% loading is considered as pseudo data. All three methods (WLS, CS, and CST) are employed to estimate the states of the 615 node network in three time steps. Table 6 compares the accuracy and computational time of these three methods. The computational time of WLS in this case study is 1.64s, while for CST algorithm with three measured data is 1.17s, as represented in Table 6. Although the simulation time for CST is more than CS, the AMVE and AAVE are significantly improved. The AMVE of CS algorithm in time step 3 is 2.5%, while by considering the impact of spatial-temporal correlation in CST, the AMVE of the estimated states decreased by 50% and reached to 1.2%. Based on the provided results in Tables 6, in terms of AMVE, AAVE, Quality, and computational time, the effectiveness of the developed DSSE spatial-temporal algorithm is clear in the large distribution networks.

### 5.4 Case Study 4: Australian LV Distribution Network with Unbalanced Loads

In the fourth case study, the proposed method is applied to an Australian residential, three phase unbalanced LV network with 23 buses, as shown in Figure 12. The node buses in this network have been selected to link with and represent several RC loads, based on the discussion provided in Section 2. The performances of CS and proposed CST estimation algorithms are compared and the enhanced efficiency of CST over CS is supported by simulation results. The residential load and PV data used for this case study are collected from real measurements located in Newmarket suburb in Brisbane, Australia.

Table 6. Comparison results for case study 3, with decreasing load

| | **Time step 1: 80% Loading** | | | **Time step 2: 60% Loading** | | | **Time step 3: 40% Loading** | | |
|---|---|---|---|---|---|---|---|---|---|
| | WLS | CS | CST | WLS | CS | CST | WLS | CS | CST |
| **AMVE (%)** | 4.9 | 0.8 | 0.8 | 5.8 | 1.3 | 0.9 | 6.9 | 2.5 | 1.2 |
| **AAVE (deg)** | 2.25 | 0.31 | 0.31 | 3.67 | 1.03 | 0.94 | 4.16 | 1.51 | 1.04 |
| **Quality** | 4.18 | 6.74 | 6.74 | 4.19 | 6.74 | 6.74 | 4.18 | 6.74 | 6.74 |
| **Time(s)** | 1.64 | 0.28 | 0.28 | 1.64 | 0.28 | 0.73 | 1.64 | 0.28 | 1.17 |
| **Iterations** | 5 | 1 | 1 | 5 | 1 | 1 | 5 | 1 | 1 |

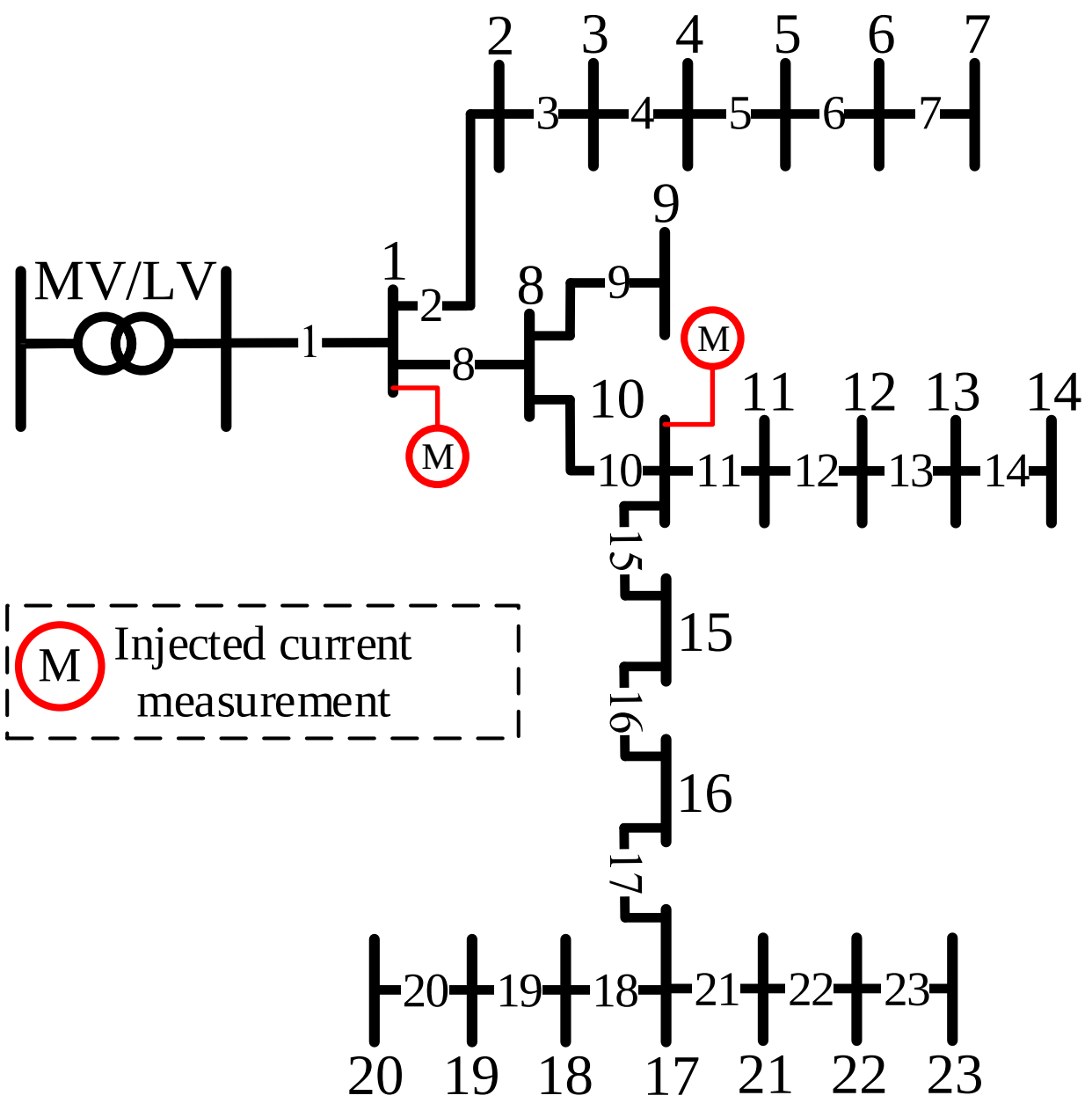

Figure 12. Schematic of an Australian residential distribution network.

The correlation matrix is calculated based on the historical data of this suburb. However, for new-built areas or those without historical data, the correlation matrix from areas with the same load type (as shown in Section 2 can be considered in DSSE formulation. The CS and CST methods are applied to the unbalanced LV network, to estimate each phase considering the mutual inductances between conductors. In this network, only bus numbers of 1 and 10 have injected current measurements. In CST algorithm, a rolling window of last three time steps of the measured data is used in (13) to increase the corrective terms of estimation based on the spatial-temporal correlation. In determining the length of the rolling window, both the accuracy improvement and computational cost should be taken into account. Including more time steps increases the computational cost and data requirement. In addition, as we move back in time the temporal correlations decline. The voltage magnitude of bus 8 and 23 are visualized in Figure 13 and 14. Bus 8 is chosen because it plays the main role in estimating the states of fourteen downstream buses. Table 7 shows the impact of temporal correlation, revealing that considering temporal correlation can consistently decrease AMVE, AAVE, maximum magnitude voltage error (MMVE) and maximum angle voltage error (MAVE).

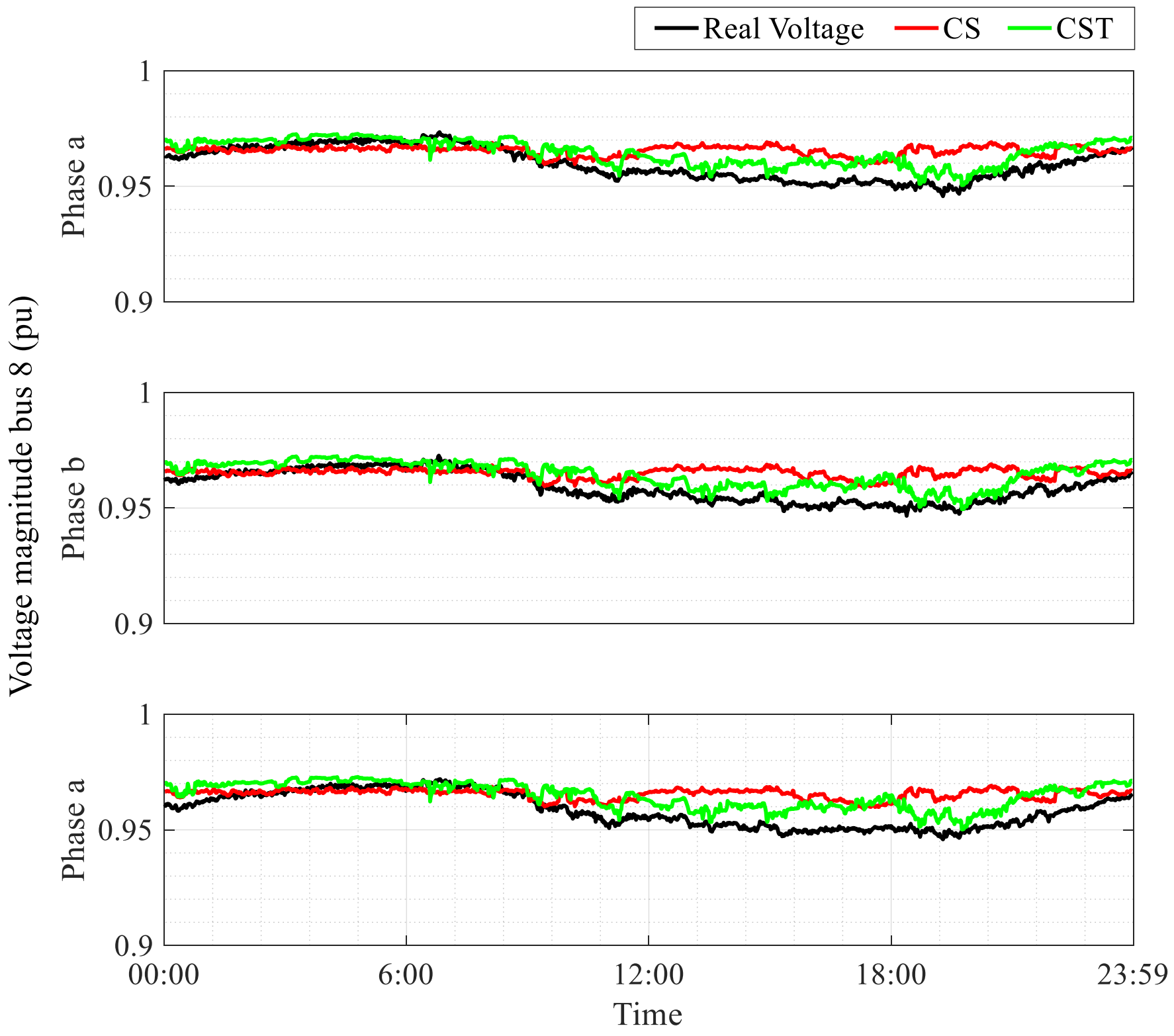

Figure 13. Three phase voltage magnitudes profile at bus 8.

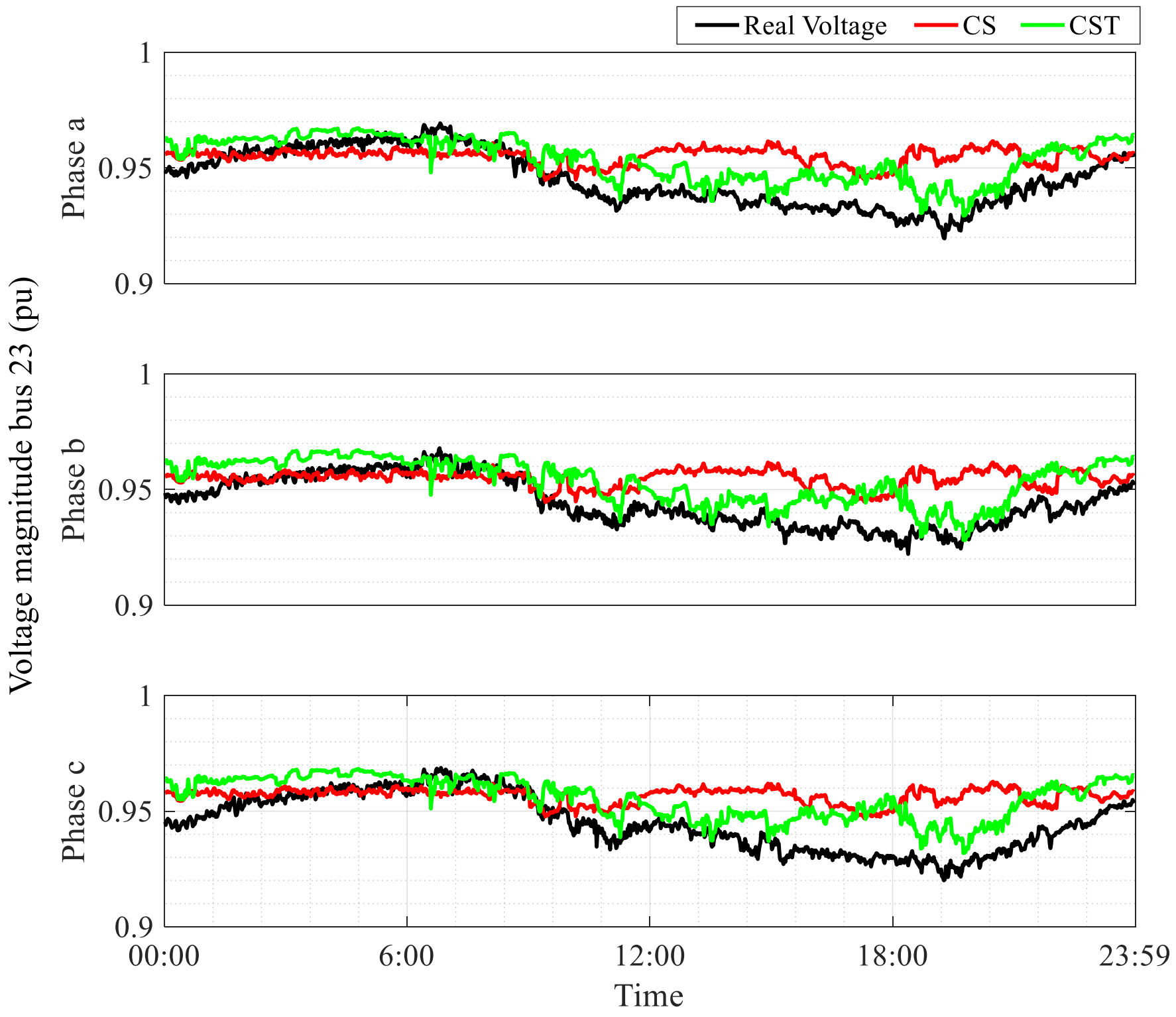

Figure 14. Three phase voltage magnitudes profile at bus 23.

Table 7. Voltage error in case study 4

| | CS | | | | CST | | | |
|---|---|---|---|---|---|---|---|---|
| | **Magnitude** | | **Phase** | | **Magnitude** | | **Phase** | |
| | AMVE (%) | MMVE (%) | AAVE (deg) | MAVE (deg) | AMVE (%) | MMVE (%) | AAVE (deg) | MAVE (deg) |
| **Bus 8** | 0.6 | 1.9 | 0.17 | 0.60 | 0.45 | 1.3 | 0.12 | 0.43 |
| **Bus 23** | 0.69 | 2 | 0.19 | 0.80 | 0.54 | 1.4 | 0.14 | 0.50 |

The average and maximum magnitude voltage errors are decreased by almost 30% in CST comparing to CS. Similarly, AMVE and MMVE of the phase voltages are improved by 30%. Bus 23 is far from the MV/LV transformer and considered as the worst impacted scenario in this case study. However, still, the average and maximum voltage errors for both magnitude and phase decrease considerably by relying on spatial-temporal correlations comparing to spatial correlation only. CST comparing to CS decreases AMVE and MMVE for the voltage magnitude by 22% and 30%, respectively. It also decreases the AAVE and MAVE of phase by 0.05 and 0.3 degrees, respectively.

To evaluate the performance of DSSE algorithms, it is important to assess their performance in the distribution networks with high R/X ratio cables [5]. Therefore, in the last part of this section, we increased the resistance of the cables for 30 %, while decreasing the inductance of the cables to keep the impedance of the cables constant, for comparison purposes. Table 8 represents the unique performance of the developed method, while the error of the estimator is slightly increased. The AMVE of the CST is slightly increased to 0.59% in presence of the higher R/X ratio in the distribution cables. It is worth to note that the angle errors reported in Table 8 are almost the same as the angle errors in Table 7.

Table 8. Voltage error in case study 4 with higher R/X ratio

| | CS | | | | CST | | | |
|---|---|---|---|---|---|---|---|---|
| | **Magnitude** | | **Phase** | | **Magnitude** | | **Phase** | |
| | AMVE (%) | MMVE (%) | AAVE (deg) | MAVE (deg) | AMVE (%) | MMVE (%) | AAVE (deg) | MAVE (deg) |
| **Bus 8** | 0.83 | 2.4 | 0.21 | 0.62 | 0.59 | 1.7 | 0.19 | 0.45 |
| **Bus 23** | 0.89 | 2.6 | 0.24 | 0.91 | 0.69 | 1.9 | 0.22 | 0.67 |

## 6. Conclusion

This paper proposes an efficient DSSE method to estimate node voltages, branch currents, and injection currents for both MV and in particular LV distribution networks with renewable resources and unbalance loads. Our investigations confirmed the significant role of deploying spatial-temporal customer loads correlation in improving the accuracy of DSSE. In order to verify the accuracy and effectiveness of the proposed method in static and real-time state estimations, four different systems under various operating scenarios are studied. An aggregation of customer loads in each RC is considered to gain higher correlation coefficients leading to higher state estimation accuracy. According to simulation results, deploying spatial-temporal customer loads correlation in DSSE improves the quality, accuracy and computational time in all case studies. The proposed scheme works efficiently in real-time simulation. Incorporation of spatial-temporal correlation compared with only spatial correlation, decreases the estimation errors by 30% in case study 3, similar improvements are reported in the case studies 1 and 2.

## References

[1] F. C. Schweppe and J. Wildes, "Power system static-state estimation, Part I: Exact model," *IEEE Transactions on Power Apparatus and Systems*, no. 1, pp. 120-125, 1970.

[2] A. S. Meliopoulos *et al.*, "Smart grid technologies for autonomous operation and control," *IEEE Transactions on Smart Grid*, vol. 2, no. 1, pp. 1-10, 2011.

[3] W.-M. Lin and J.-H. Teng, "Distribution fast decoupled state estimation by measurement pairing," *IEE Proceedings-Generation, Transmission and Distribution*, vol. 143, no. 1, pp. 43-48, 1996.

[4] Y. Zhang, X. Du, and M. Salman, "Battery state estimation with a self-evolving electrochemical ageing model," *International Journal of Electrical Power & Energy Systems*, vol. 85, pp. 178-189, 2017.

[5] A. Primadianto and C.-N. Lu, "A review on distribution system state estimation," *IEEE Transactions on Power Systems*, vol. 32, no. 5, pp. 3875-3883, 2017.

[6] I. Džafić, R. A. Jabr, I. Huseinagić, and B. C. Pal, "Multi-phase state estimation featuring industrial-grade distribution network models," *IEEE Transactions on Smart Grid*, vol. 8, no. 2, pp. 609-618, 2017.

[7] A. Arefi and M.-R. Haghifam, "State estimation in smart power grids," in *Smart Power Grids 2011*: Springer, 2012, pp. 439-478.

[8] A. Arefi, M. R. Haghifam, and S. H. Fathi, "Distribution harmonic state estimation based on a modified PSO considering parameters uncertainty," in *PowerTech, 2011 IEEE Trondheim*, pp. 1-7, 2011.

[9] X. Qing, H. R. Karimi, Y. Niu, and X. Wang, "Decentralized unscented Kalman filter based on a consensus algorithm for multi-area dynamic state estimation in power systems," *International Journal of Electrical Power & Energy Systems,* vol. 65, pp. 26-33, 2015.
[10] O. Krause, D. Martin, and S. Lehnhoff, "Under-determined WLMS state estimation," in *Power and Energy Engineering Conference (APPEEC), 2015 IEEE PES Asia-Pacific,* pp. 1-6, 2015.
[11] M. E. Baran and A. W. Kelley, "State estimation for real-time monitoring of distribution systems," *IEEE Transactions on Power Systems,* vol. 9, no. 3, pp. 1601-1609, 1994.
[12] A. K. Ghosh, D. L. Lubkeman, M. J. Downey, and R. H. Jones, "Distribution circuit state estimation using a probabilistic approach," *IEEE Transactions on Power Systems,* vol. 12, no. 1, pp. 45-51, 1997.
[13] I. Roytelman and S. Shahidehpour, "State estimation for electric power distribution systems in quasi real-time conditions," *IEEE Transactions on Power Delivery,* vol. 8, no. 4, pp. 2009-2015, 1993.
[14] C. Muscas, M. Pau, P. A. Pegoraro, and S. Sulis, "Effects of measurements and pseudomeasurements correlation in distribution system state estimation," *IEEE Transactions on Instrumentation and Measurement,* vol. 63, no. 12, pp. 2813-2823, 2014.
[15] G. Valverde, A. T. Saric, and V. Terzija, "Stochastic monitoring of distribution networks including correlated input variables," *IEEE Transactions on power delivery,* vol. 28, no. 1, pp. 246-255, 2013.
[16] Y. R. Gahrooei, A. Khodabakhshian, and R.-A. Hooshmand, "A New Pseudo Load Profile Determination Approach in Low Voltage Distribution Networks," *IEEE Transactions on Power Systems,* vol. 33, no. 1, pp. 463-472, 2018.
[17] F. Golestaneh, H. B. Gooi, and P. Pinson, "Generation and evaluation of space–time trajectories of photovoltaic power," *Applied Energy,* vol. 176, pp. 80-91, 2016.
[18] J. D. Melo, E. M. Carreno, and A. Padilha-Feltrin, "Estimation of a preference map of new consumers for spatial load forecasting simulation methods using a spatial analysis of points," *International Journal of Electrical Power & Energy Systems,* vol. 67, pp. 299-305, 2015.
[19] Y. Chakhchoukh, V. Vittal, and G. T. Heydt, "PMU based state estimation by integrating correlation," *IEEE Transactions on Power Systems,* vol. 29, no. 2, pp. 617-626, 2014.
[20] S. S. Alam, B. Natarajan, and A. Pahwa, "Distribution grid state estimation from compressed measurements," *IEEE Transactions on Smart Grid,* vol. 5, no. 4, pp. 1631-1642, 2014.
[21] J. F. Buford, X. Wu, and V. Krishnaswamy, "Spatial-temporal event correlation," in. *IEEE International Conference on Communications ICC'09*, pp. 1-6, 2009.
[22] J. Lee Rodgers and W. A. Nicewander, "Thirteen ways to look at the correlation coefficient," *The American Statistician,* vol. 42, no. 1, pp. 59-66, 1988.
[23] P. Street, "Dataport: the world's largest energy data resource," *Pecan Street Inc,* 2015.
[24] S. Kanna, D. H. Dini, Y. Xia, S. Hui, and D. P. Mandic, "Distributed widely linear Kalman filtering for frequency estimation in power networks," *IEEE Transactions on Signal and Information Processing over Networks,* vol. 1, no. 1, pp. 45-57, 2015.
[25] B. Picinbono, "Second-order complex random vectors and normal distributions," *IEEE Transactions on Signal Processing,* vol. 44, no. 10, pp. 2637-2640, 1996.
[26] R. Singh, B. Pal, and R. Jabr, "Distribution system state estimation through Gaussian mixture model of the load as pseudo-measurement," *IET generation, transmission & distribution,* vol. 4, no. 1, pp. 50-59, 2010.
[27] A. Abur and A. G. Exposito, *Power system state estimation: theory and implementation*. CRC press, 2004.
[28] A. Arefi, M. R. Haghifam, and S. H. Fathi, "Observability analysis of electric networks considering branch impedance," *International Journal of Electrical Power & Energy Systems,* vol. 33, no. 4, pp. 954-960, 2011.

[29] J.-H. Teng, "A direct approach for distribution system load flow solutions," *IEEE Transactions on Power Delivery*, vol. 18, no. 3, pp. 882-887, 2003.
[30] D. A. Haughton and G. T. Heydt, "A linear state estimation formulation for smart distribution systems," *IEEE Transactions on Power Systems,* vol. 28, no. 2, pp. 1187-1195, 2013.
[31] C. Muscas, F. Pilo, G. Pisano, and S. Sulis, "Optimal allocation of multichannel measurement devices for distribution state estimation," *IEEE Transactions on Instrumentation and Measurement*, vol. 58, no. 6, pp. 1929-1937, 2009.
[32] P. J. Schreier and L. L. Scharf, *Statistical signal processing of complex-valued data: the theory of improper and noncircular signals*. Cambridge University Press, 2010.
[33] S. L. Goh and D. P. Mandic, "An augmented extended Kalman filter algorithm for complex-valued recurrent neural networks," *Neural Computation*, vol. 19, no. 4, pp. 1039-1055, 2007.
[34] D. Bates and M. Maechler, "R package Matrix: Sparse and dense matrix classes and methods," 2013.
[35] Z. Liu, F. Wen, and G. Ledwich, "Optimal planning of electric-vehicle charging stations in distribution systems," *IEEE Transactions on Power Delivery*, vol. 28, no. 1, pp. 102-110, 2013.
[36] T. Yamane, "Statistics: An introductory analysis," 1973.
[37] D. T. Feeders, "IEEE PES Distribution System Analysis Subcommittee's, Distribution Test Feeder Working Group," 2013.